\documentclass[aps,prb,citeautoscript,twocolumn,longbibliography,superscriptaddress]{revtex4-1}
\usepackage[T1]{fontenc}
\usepackage{latexsym}
\usepackage{graphicx} 
\usepackage{epstopdf}
\usepackage{amsmath}  
\usepackage{amssymb}
\usepackage{comment}
\usepackage{cases}
\usepackage{lipsum}
\usepackage{float}
\usepackage{tikz}
\usepackage{makecell}
\usepackage{bbold}

\usepackage{simplewick}
\usepackage{soul}
\usepackage{color}

\usepackage[breaklinks=true]{hyperref}

\begin{document}

\title{Collective spin and charge excitations in the $t$-$J$-$U$ model of high-$T_c$ cuprates}

\author{Maciej Fidrysiak}
\email{maciej.fidrysiak@uj.edu.pl}
\affiliation{Institute of Theoretical Physics, Jagiellonian University, ul. {\L}ojasiewicza 11, 30-348 Krak{\'o}w, Poland }
\author{Danuta Goc-Jag{\l}o}
\email{danuta.goc-jaglo@uj.edu.pl}
\affiliation{Institute of Theoretical Physics, Jagiellonian University, ul. {\L}ojasiewicza 11, 30-348 Krak{\'o}w, Poland }
\author{J{\'o}zef Spa{\l}ek}%
\email{jozef.spalek@uj.edu.pl}
\affiliation{Institute of Theoretical Physics, Jagiellonian University, ul. {\L}ojasiewicza 11, 30-348 Krak{\'o}w, Poland }

\begin{abstract}
  The $t$-$J$-$U$ model of high-$T_c$ copper-oxide superconductors incorporates both the on-site Coulomb repulsion and kinetic exchange interaction and yields a semi-quantitative description of the static properties of those materials. We extend this analysis to dynamic quantities and address collective spin- and charge excitations in the correlated metallic state of the $t$-$J$-$U$ model. We employ VWF+$1/\mathcal{N}_f$ approach that combines the variational wave function (VWF) approach with the expansion in the inverse number of fermionic flavors ($1/\mathcal{N}_f$). It is shown that the resonant (paramagnon) contribution to the dynamic magnetic susceptibility remains robust as one interpolates between the Hubbard- and $t$-$J$-model limits, whereas the incoherent continuum undergoes substantial renormalization. Energy of the collective charge mode diminishes as the strong-coupling limit is approached. We also introduce the concept of effective kinetic exchange interaction that allows for a unified interpretation of magnetic dynamics in the Hubbard, $t$-$J$, and $t$-$J$-$U$ models. The results are discussed in the context of recent resonant inelastic $x$-ray scattering experiments for the high-$T_c$ cuprates.
\end{abstract}

\maketitle

\section{Introduction}
\label{sec:introduction}

High-temperature (high-$T_c$) copper-oxide superconductors (SC) serve as paradigmatic strongly-correlated systems, evolving from the antiferromagnetic (AF) Mott insulating state, through high-$T_c$ SC state, to normal metal phase as a function of chemical doping. The common theoretical frameworks, used to describe them, are based on (extended) Hubbard and $t$-$J$-models, yet a unified quantitative description of static- and single-particle properties of high-$T_c$ cuprates for a fixed set of microscopic parameters has not been achieved so far within those schemes. In effect, extended $t$-$J$-$U$ Hamiltonian, encompassing both Hubbard and $t$-$J$ models as special cases, has been proposed and demonstrated to yield a consistent description of principal experimental data for high-$T_c$ cuprates within variational scheme \cite{SpalekPhysRevB2017,ZegrodnikPhysRevB2017,ZegrodnikPhysRevB2017_2,ZegrodnikPhysRevB2018}. However, the $t$-$J$-$U$-model studies have been hitherto restricted to static properties, wheres recent developments in spectroscopic techniques (in particular, resonant inelastic $x$-ray scattering (RIXS)) provide comprehensive evidence for the relevance of collective excitations across the phase diagram of high-$T_c$ cuprates.\cite{DeanNatMater2013,IshiiNatCommun2014,LeeNatPhys2014,GuariseNatCommun2014,WakimotoPhysRevB2015,MinolaPhysRevLett2017,IvashkoPhysRevB2017,MeyersPhysRevB2017,ChaixPhysRevB2018,Robarts_arXiV_2019,ZhouNatCommun2013,GretarssonPhysRevLett2016,FumagalliPhysRevB2019,LeTaconNatPhys2011,JiaNatCommun2014,PengPhysRevB2018,IshiiPhysRevB2017,HeptingNature2018,IshiiJPhysSocJapan2019,LinNPJQuantMater2020,SinghArXiV2020,NagPhysRevLett2020} In this respect, the $t$-$J$-$U$ model remains largely unexplored and is yet to be tested as a tool to study many-particle dynamics. 

Here we analyze the structure of dynamical spin and charge susceptibilities in the paramagnetic metallic state of the two-dimensional $t$-$J$-$U$ model and relate its predictions to those of the Hubbard and $t$-$J$ models that are used extensively \cite{OvchinnikovBook2004,FidrysiakPhysRevB2020,FidrysiakPhysRevB2021,FidrysiakArXiV2021,GrecoPhysRevB2016,GrecoJPSJ2017,GrecoPhysRevB2020} to study collective modes in correlated electron systems. We employ VWF+$1/\mathcal{N}_f$ scheme \cite{FidrysiakPhysRevB2020,FidrysiakPhysRevB2021,FidrysiakArXiV2021}, combining \textbf{V}ariational \textbf{W}ave \textbf{F}unction (VWF) approach in its diagrammatic form with the field-theoretical $1/\mathcal{N}_f$ expansion ($\mathcal{N}_f$ denotes the number of fermionic flavors). The latter has been recently benchmarked\cite{FidrysiakPhysRevB2021} against determinant quantum Monte-Carlo for small systems. This allows us to incorporate both the effect of local correlations and long-range collective excitations, required to be able to analyze  the strong-coupling ($t$-$J$-model) limit. We find that the resonant part of the magnetic response (paramagnon mode) remains comparable within Hubbard, $t$-$J$, and $t$-$J$-$U$ models, provided that the parameters of those three Hamiltonians are appropriately mapped onto each other. On the other hand, the incoherent part of the spectrum is substantially affected by variation of the Hubbard $U$. We also find that the charge mode energy undergoes reduction as one moves from the Hubbard to the $t$-$J$-model regime, reflecting enhanced band renormalization effects near the $t$-$J$-model limit. This analysis supports the $t$-$J$-$U$ model as a tool to study semi-quantitatively collective modes in high-$T_c$ cuprates.

\begin{figure*}
  \centering
  \includegraphics[width=1\linewidth]{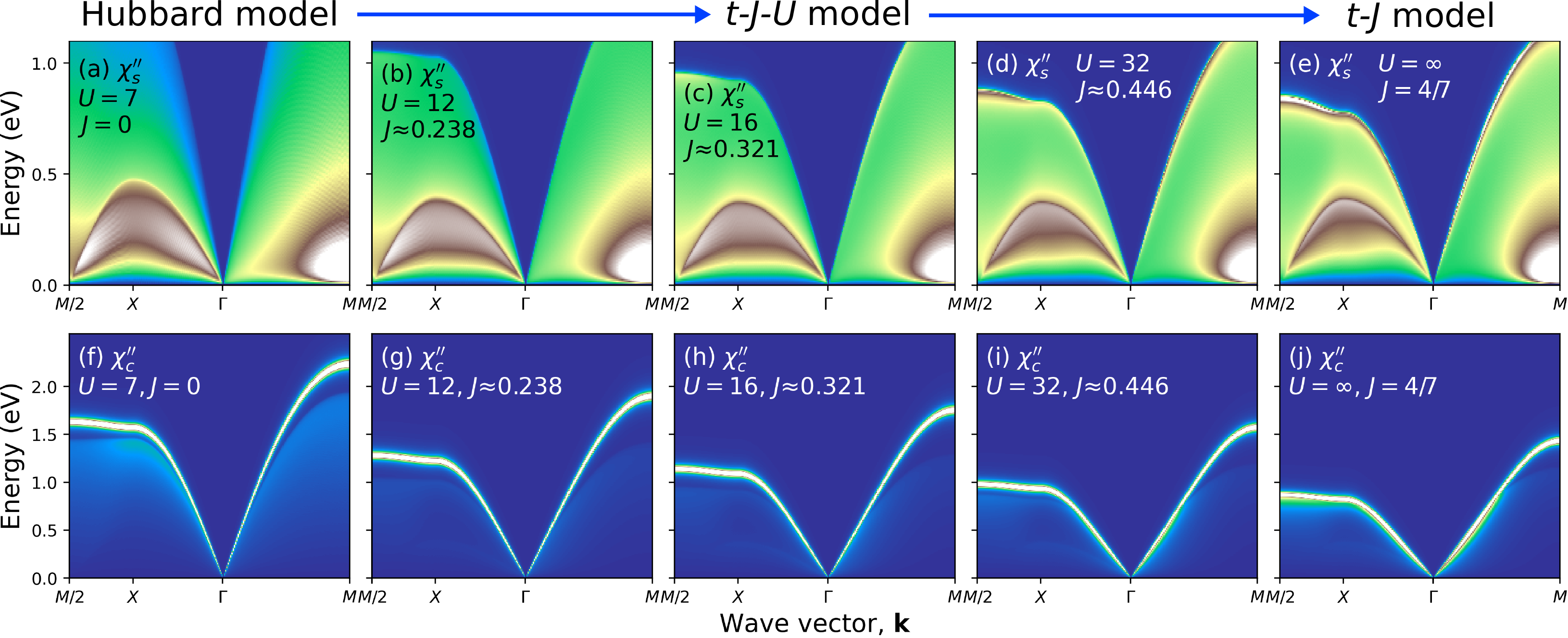}
  \caption{Calculated imaginary parts of spin (top panels) and charge (bottom panels) dynamical susceptibilities for the $t$-$J$-$U$ model along the $M/2$-$X$-$\Gamma$-$M$ contour. The common value of effective exchange interaction $J_\mathrm{eff} = J +\frac{4t^2}{U} \equiv 0.2\,\mathrm{eV}$ is adopted for all panels, whereas the on-site repulsion $U$ varies from $7$ on the left [(a) and (f)] to $\infty$ on the right [(e) and (j)]. The parameters are detailed inside the panels in units of $|t|$. Blue and white colors map to low- and high intensity, respectively. Note the model evolution marked on the top.}
  \label{fig:color_map}
\end{figure*}

\section{Model and method}
\label{sec:model}

We employ the $t$-$J$-$U$ model on two-dimensional $200 \times 200$ square lattice, given by the Hamiltonian

\begin{align}
  \label{eq:tju-model}
  \hat{\mathcal{H}} = \sum_{i\neq j, \sigma} t_{ij} \hat{c}_{i\sigma}^\dagger \hat{c}_{j\sigma} + U \sum_i \hat{n}_{i\uparrow} \hat{n}_{i\downarrow} + J \sum_{\langle i, j\rangle} \hat{\mathbf{S}}_i \hat{\mathbf{S}}_j,
\end{align}

\noindent
where $\hat{c}_{i\sigma}$ ($\hat{c}_{i\sigma}^\dagger$) denote electron annihilation (creation) operators on site $i$  with spin index $\sigma = \pm 1$, $\hat{n}_{i\sigma} \equiv \hat{c}_{i\sigma}^\dagger \hat{c}_{i\sigma}$, and $\hat{\mathbf{S}}_i \equiv (\hat{S}^x_i, \hat{S}^y_i, \hat{S}^z_i)$ are spin operators. We retain only nearest- and next-nearest hopping integrals, $t \equiv - 0.35\,\mathrm{eV}$ and $t^\prime \equiv 0.25 |t|$, respectively. The on-site Coulomb repulsion, $U$, and exchange coupling, $J$, are treated as free parameters. To interpret the results on the $U$-$J$ plane, we introduce \emph{effective kinetic exchange interaction}, $J_\mathrm{eff} \equiv J + \frac{4t^2}{U}$, combining $J$ with the second-order kinetic exchange that results from the canonical perturbation expansion\cite{ChaoPhysRevB1978} for the Hubbard model. Unless stated otherwise, we restrict the parameter space by setting  $J_\mathrm{eff} \equiv 0.2\,\mathrm{eV}$ so that the exchange coupling is determined by $U \ge 7 |t|$ as $J = J_{\mathrm{eff}} - \frac{4t^2}{U}$. In the Hubbard model limit ($J = 0$, $U = 7\,|t|$), this choice yields a semi-quantitative agreement with measured paramagnon spectra for the lanthanum cuprates \cite{FidrysiakPhysRevB2020}. Moreover, the selection of sizable $J_\mathrm{eff}$ favors AF correlations and thus places the system away from ferromagnetism \cite{SpalekPhysstatSol1981,EisenbergPhysRevB2002,LiuPhysRevLett2012,BlesioPhysRevB2019} (cf. the phase stability analysis below). Hereafter we set the hole concentration to $\delta = 0.16$ and temperature $k_B T = 0.4 |t|$ to ensure stability of the paramagnetic state against fluctuations.

The model is solved using VWF+$1/\mathcal{N}_f$ approach in the local-diagrammatic (LD) variant,  $\mathrm{LD}_f+1/\mathcal{N}_f$.\cite{FidrysiakPhysRevB2021} First, the energy functional $E_\mathrm{var} \equiv \langle \Psi_\mathrm{var}| \hat{\mathcal{H}} |\Psi_\mathrm{var}\rangle$ is constructed based on the $t$-$J$-$U$ Hamiltonian~\eqref{eq:tju-model} and the trial state, $|\Psi_\mathrm{var}\rangle \equiv C \prod_i\hat{P}_i |\Psi_0\rangle$, where $|\Psi_0\rangle$ represents Slater determinant to be determined self-consistently,

\begin{align}
\hat{P}_i  \equiv \lambda_0|0\rangle_{ii}\langle0| + \sum_{\sigma\sigma^\prime}\lambda_{\sigma\sigma^\prime} |\sigma\rangle_{ii}\langle\sigma^\prime| + \lambda_{d} |{\uparrow\downarrow}\rangle_{ii}\langle{\uparrow\downarrow}|
\end{align}

\noindent
introduces local correlations into $|\Psi_\mathrm{var}\rangle$, and $C$ is the normalization factor. The correlators, $\hat{P}_i$, are expressed in terms of the local many-particle basis on site $i$, encompassing empty ($|0\rangle_i$), singly-occupied ($|{\uparrow}\rangle_i$, $|{\downarrow}\rangle_i$), and doubly occupied ($|{\uparrow\downarrow}\rangle_i$) configurations. The six coefficients, $\lambda_0$, $\lambda_{\sigma\sigma\prime}$ ($\sigma, \sigma^\prime = \uparrow, \downarrow$), and $\lambda_d$, serve as variational parameters. By application Wick's theorem, energy is then expressed as $E_\mathrm{var} = E_\mathrm{var}(\mathbf{P}, \boldsymbol{\lambda})$, where $\mathbf{P}$ is a vector composed of two-point correlation functions (\emph{lines}) in the form $P_{i\sigma, j\sigma^\prime} = \langle\Psi_0|\hat{c}^\dagger_{i\sigma} \hat{c}_{j\sigma^\prime}|\Psi_0\rangle$, and $\boldsymbol{\lambda}$ denotes correlator parameters. We evaluate $E_\mathrm{var}$ within specialized diagrammatic scheme\cite{BunemannEPL2012,KaczmarczykNewJPhys2014} and retain only local diagrams, which results in the so-called local diagrammatic (LD) approximation \cite{FidrysiakPhysRevB2021}. The latter incorporates multiple-loop diagrams and goes beyond the renormalized mean-field theory already at the zeroth-order (saddle-point) level. Both $\mathbf{P}$ and $\boldsymbol{\lambda}$ are then treated as (imaginary-time) dynamical fields, and their quantum and classical fluctuations around this correlated state are studied at the leading order in $1/\mathcal{N}_f$ expansion (note that the fields are subjected to additional constraints, cf. Ref.~\citenum{FidrysiakPhysRevB2021}). This allows us to calculate imaginary-time dynamical spin and charge susceptibilities ($\chi_s$ and $\chi_c$, respectively), which are analytically continued to real frequencies as $i\omega_n \rightarrow \omega + i 0.02 |t|$.

\section{Results}
\label{sec:results}

In Fig.~\ref{fig:color_map}, the calculated imaginary parts of spin (top) and charge (bottom) dynamical susceptibilities for the $t$-$J$-$U$ model~\eqref{eq:tju-model} with fixed $J_\mathrm{eff} \equiv 0.2\,\mathrm{eV}$, are displayed. The left panels, (a) and (f), correspond to the Hubbard model ($U = 7 |t|$ and $J = 0$), whereas (e) and (j) represent the $t$-$J$-model limit ($U = \infty$, $J = \frac{4}{7} |t|$). The middle panels show the results for intermediate values of the Hubbard $U$ (parameters are detailed inside the figure). As is apparent from Fig.~\ref{fig:color_map}(a)-(e), the magnetic spectrum separates into intense and dispersive paramagnon contribution (ranging up to $\sim 0.3\,\mathrm{eV}$) and the incoherent part. The latter may be interpreted as particle-hole continuum, which has been verified explicitly in the Hubbard-model limit by comparison with correlated Lindhard susceptibility.\cite{FidrysiakArXiV2021} Remarkably, well-defined paramagnon persists along the anti-nodal ($\Gamma$-$X$) line and the magnetic Brillouin zone boundary ($X$-$M/2$) in the metallic state ($\delta = 0.16$), but it is absent in the nodal ($\Gamma$-$M$) direction. This shows that robust magnetic excitations may arise in strongly correlated systems and resist overdamping, even without long-range magnetic order. In the Hubbard-model limit, those highly anisotropic paramagnon characteristics have been recently shown to semi-quantitatively reproduce experimental inelastic neutron scattering and RIXS data for selected high-$T_c$ superconductors \cite{FidrysiakPhysRevB2020,FidrysiakArXiV2021}. As the main finding of the present study, we demonstrate  \emph{weak dependence of the peak paramagnon energy on $U$} for a fixed value of  $J_\mathrm{eff}$ [cf. Fig.~\ref{fig:color_map}(a)-(e)]. This supports the proposed here interpretation of $J_\mathrm{eff}$ as the scale controlling paramagnons in the metallic state of the $t$-$J$-$U$ model close to AF instability and should allow us to reconcile the Hubbard-model results for dynamic quantities with the $t$-$J$-$U$-($V$) model analysis\cite{SpalekPhysRevB2017,ZegrodnikPhysRevB2017,ZegrodnikPhysRevB2017_2,ZegrodnikPhysRevB2018} of equilibrium properties. The particle-hole excitations exhibit qualitatively different behavior and are shifted to lower energies as the on-site Coulomb interaction increases, which can be attributed to correlation-induced narrowing of single-particle bandwidth. The coherent- and incoherent magnetic excitations are thus, to large extent, decoupled and governed by distinct energy scales ($J_\mathrm{eff}$ and renormalized bandwidth, respectively).

\begin{figure}
  \centering
  \includegraphics[width=0.98\linewidth]{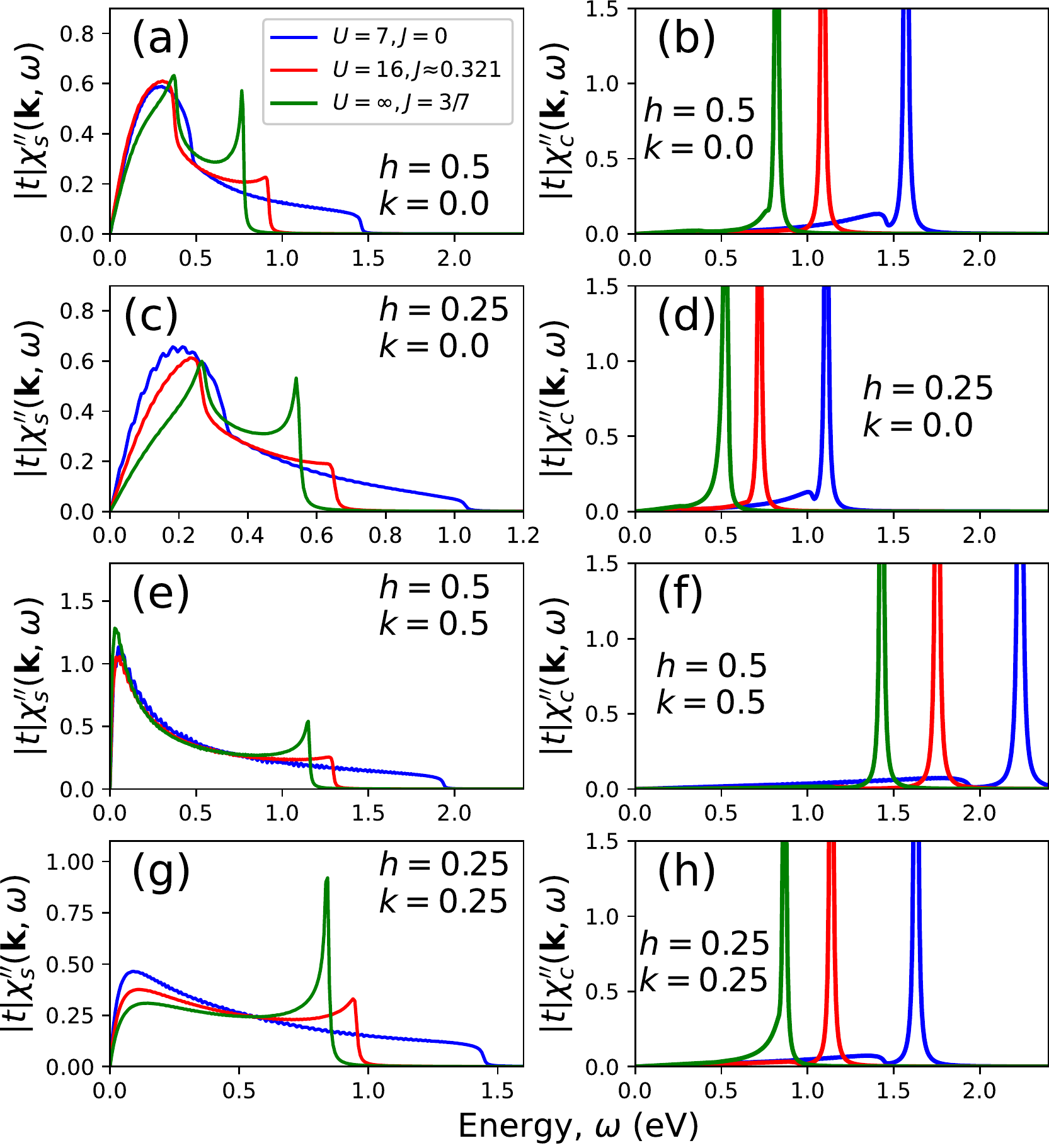}
  \caption{Calculated energy dependence of the imaginary parts of spin [(a), (c), (e), and (g)] and charge [(b), (d), (f), and (h)] dynamical susceptibilities for the $t$-$J$-$U$ model. Wave vectors are related to the values of $h$ and $k$ as $\mathbf{k} = (\frac{2\pi}{a} h, \frac{2\pi}{a} k)$, with $a$ being lattice constant. The common value of $J_\mathrm{eff} \equiv 0.2\,\mathrm{eV}$ is adopted for all panels. Blue-, red-, and green lines correspond to $U = 7 |t|$ (Hubbard model), $U = 16 |t|$, and $U = \infty$ ($t$-$J$ model), respectively. Energies are in units of $|t|$.}
  \label{fig:susc_cuts}
\end{figure}

\begin{figure}
  \centering
  \includegraphics[width=0.95\linewidth]{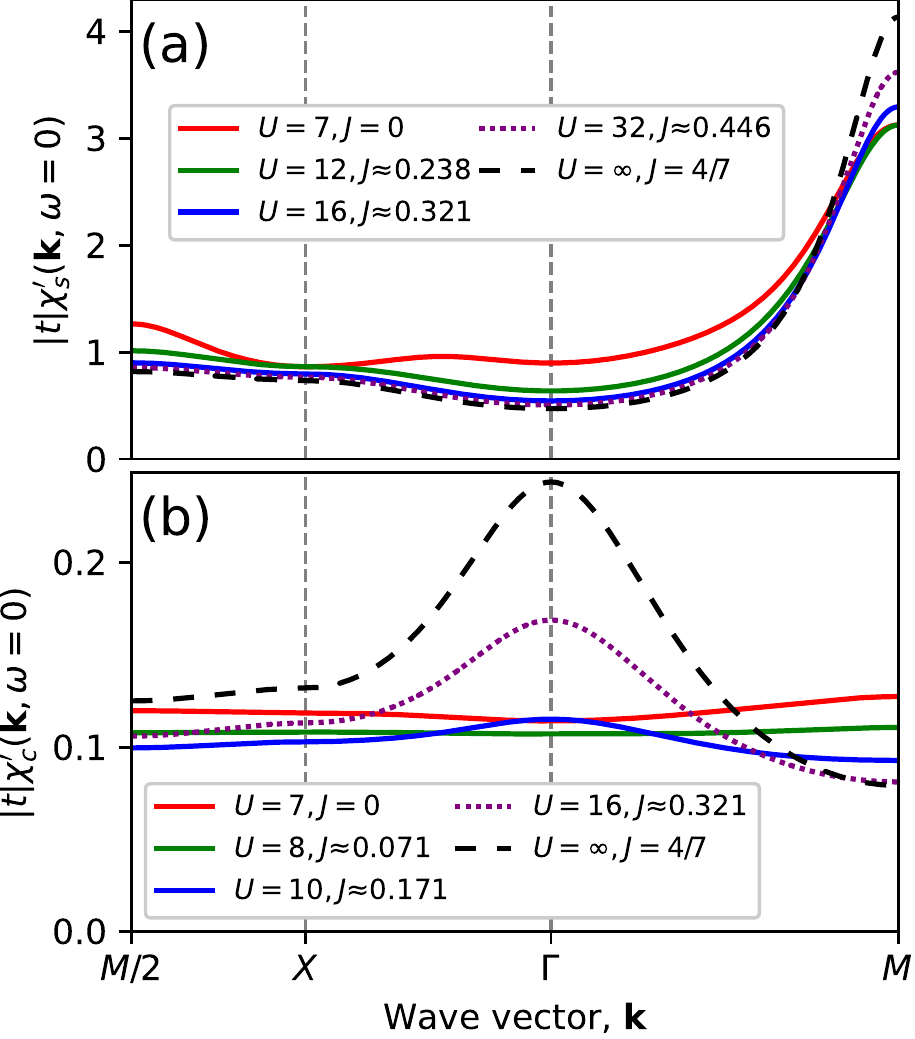}
  \caption{Static $t$-$J$-$U$-model spin (a) and charge (b) susceptibilities along the $M/2$-$X$-$\Gamma$-$M$ contour, evaluated for the same parameter range as that used to generate Figs.~\ref{fig:color_map} and \ref{fig:susc_cuts}. Here, analytic continuation has been carried out as $i \omega_n \rightarrow \omega$. The parameters for each of the curves are listed inside the panels (in units of $|t|$).}
  \label{fig:phase_stability}
\end{figure}

Charge excitations may be decomposed in an analogous manner into coherent- and incoherent components, with a well-defined mode emerging above the continuum. From Fig.~\ref{fig:color_map}(f)-(j) it follows that the charge mode energy undergoes reduction as the system evolves from the Hubbard- to the $t$-$J$-model limits. Remarkably, the charge-mode energy approaches zero near the $\Gamma$ point as a consequence of neglecting long-range three-dimensional Coulomb repulsion in the model~\eqref{eq:tju-model}, whose singular small-$\mathbf{k}$ behavior is known to open the $\Gamma$-point plasmon gap and induce dimensional crossovers in layered systems.\cite{KresinPhysRevB1988} Both effects have been observed experimentally,\cite{HeptingNature2018,LinNPJQuantMater2020,NagPhysRevLett2020} so long-range interactions are necessary to reliably describe plasmons in the cuprates.

In Fig.~\ref{fig:susc_cuts} we display the energy dependence of the calculated imaginary parts of spin (left panels) and charge (right panels) dynamical susceptibilities for the Hubbard model ($U = 7$; red lines), $t$-$J$-$U$ model ($U = 16$; blue lines), and the $t$-$J$ model ($U = \infty$; green lines), all with $J_\mathrm{eff} \equiv 0.2\,\mathrm{eV}$. The panels correspond to representative points in the Brillouin zone, detailed inside the plot (wave vectors are related to $h$ and $k$ as $\mathbf{k} = (\frac{2\pi}{a} h, \frac{2\pi}{a} k)$, with $a$ being lattice constant). Several features of the spectra may be noted. First, a well-defined peak emerges from the continuum only along the anti-nodal ($\Gamma$-$X$) line, with maximum intensity corresponding to energy  $\sim 0.3\,\mathrm{eV}$ at the $X$ point [panel (a)]. Its position weakly depends on $U$. For the nodal ($\Gamma$-$M$) direction, the spin response does not contain a clear paramagnon feature [(e) and (g)]. Second, the particle-hole continuum shifts towards lower energies for increasing $U$, with an intensity peak systematically building up at its boundary. Remarkably, this second peak appears only for $U$ substantially exceeding the bare quasiparticle bandwidth [cf. red ($U = 16 |t|$) and green ($U = \infty$) lines in Fig.~\ref{fig:susc_cuts}]. The lack of a double-peak signature in RIXS experiments supports thus the $t$-$J$-$U$ over the $t$-$J$ model as an appropriate starting point for variational studies of collective dynamics in the cuprates.

For completeness, in Fig.~\ref{fig:phase_stability} the stability of paramagnetic state against spin (a) and charge (b) fluctuations is verified for the parameter range encompassing that used in the analysis of collective excitations (cf. Figs.~\ref{fig:color_map} and \ref{fig:susc_cuts}). The static susceptibilities are displayed along the high-symmetry $M/2$-$X$-$\Gamma$-$M$ contour (all curves correspond to $J_\mathrm{eff} = 0.2\,\mathrm{eV}$ and varying values of $U$ and $J$, detailed inside the panels). Both magnetic and charge responses remain finite in the entire parameter range, implying local stability of the paramagnetic state against fluctuations. The spin susceptibility attains a maximum at the $M$ point, reflecting the dominant role of AF correlations. The charge susceptibility is peaked at the $\Gamma$ point, which is characteristic of the model with short-range interactions. The singular tail of the Coulomb repulsion tends to suppress the charge response around the $\Gamma$-point.\cite{AristovPhysRevB2006,FidrysiakArXiV2021}

\begin{figure}
  \centering
  \includegraphics[width=0.95\linewidth]{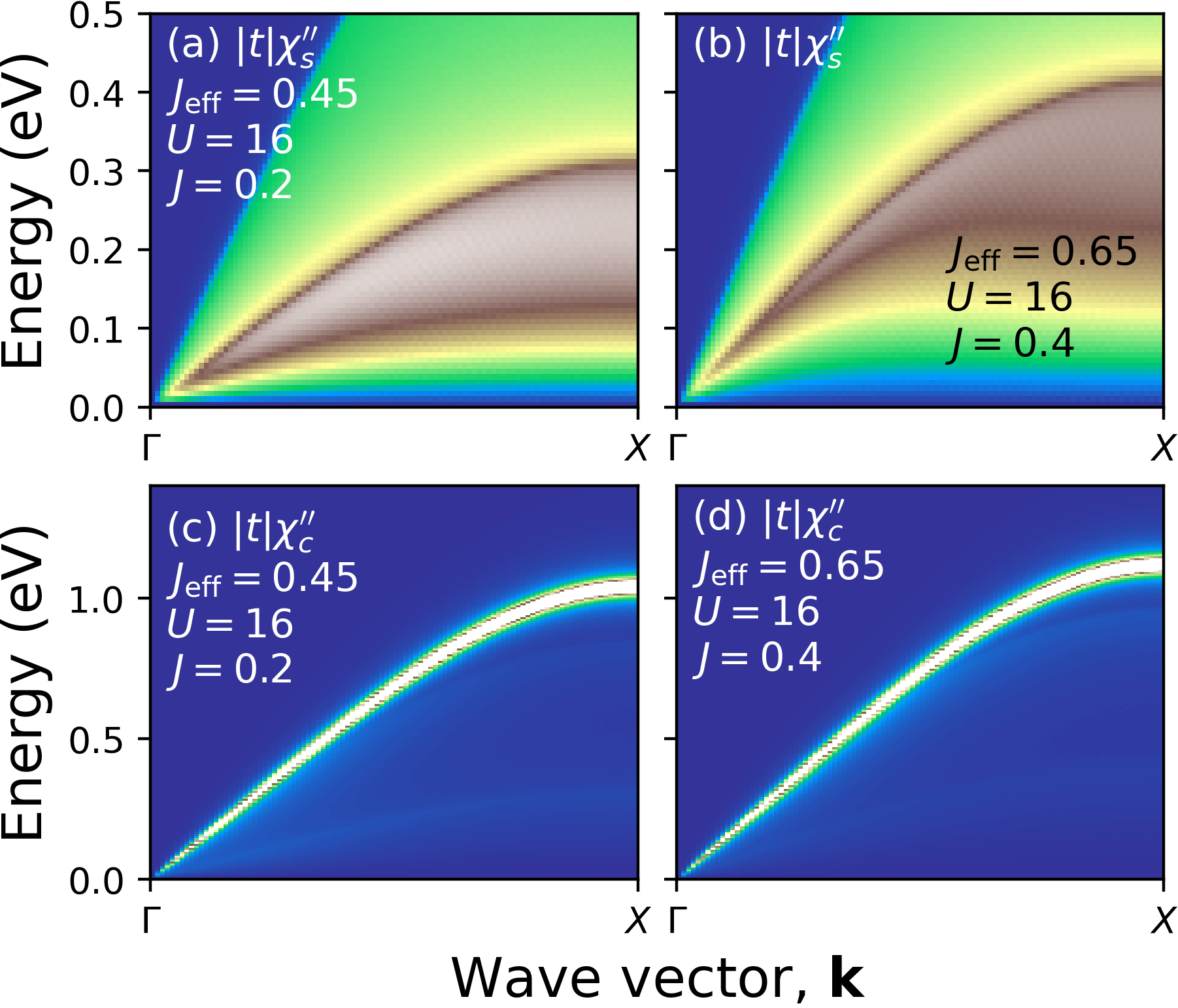}
  \caption{Imaginary parts of spin (top panels) and charge (bottom panels) dynamical susceptibilities for the $t$-$J$-$U$ model along the $M/2$-$X$-$\Gamma$-$M$ contour for $U = 16 |t|$, and two selections of $J=0.2 |t|$ [(a), (c)] and $J = 0.4 |t|$ [(b), (d)], resulting in $J_\mathrm{eff} = 0.45 |t|$ and $J_\mathrm{eff} = 0.65 |t|$, respectively. Model parameters inside the plot are given in units of $|t|$.}
  \label{fig:susc_cuts_varjeff}
\end{figure}

Finally, we examine the dependence of the paramagnon energy on $J_\mathrm{eff}$. In Fig.~\ref{fig:susc_cuts_varjeff}, we display the calculated imaginary parts of the spin (top) and charge (bottom) dynamical response for the same value of $U = 16\,|t|$, and two selections of $J = 0.2 |t|$ [(a), (c)] and $J = 0.4\,|t|$ [(b), (d)]. This results in $J_\mathrm{eff} = 0.45\,|t|$ and $J_\mathrm{eff} = 0.65\,|t|$ for left- and right panels, respectively. Fig.~\ref{fig:susc_cuts_varjeff} shows that the paramagnon energy is indeed sensitive to $J_\mathrm{eff}$ [(a)-(b)], whereas renormalization of charge mode energy due to variation of $J_\mathrm{eff}$ is less significant [(c)-(d)]. This concludes the analysis of $J_\mathrm{eff}$ as the energy scale governing paramagnon dynamics in the $t$-$J$-$U$ model with dominant AF correlations. The opposite is true for the charge excitations, which is sensitive to the value of $U$.

\section{Summary and outlook}
\label{sec:discussion}

We have analyzed the structure of spin and charge excitations in the $t$-$J$-$U$ model~\eqref{eq:tju-model} of high-$T_c$ cuprates within the VWF+$1/\mathcal{N}_f$ scheme. The effective exchange interaction, $J_\mathrm{eff} \equiv J + \frac{4t^2}{U}$, has been identified as the energy scale controlling the anti-nodal paramagnon energies close to AF ordering, which rationalizes experimentally reported similarity of $\Gamma$-$X$ magnetic excitations in metallic and AF insulating state. It also means that the paramagnon dynamics is governed predominantly by short-range spin-flip correlations (either introduced through $J$ or those originating from second-order exchange $\propto t^2/U$), the same that provide pairing potential within the local-pairing scenario of high-$T_c$ SC. Whereas generic occurrence of those magnetic excitations in the cuprates has revived the discussion about their role in high-$T_c$ SC,\cite{LeTaconNatPhys2011,DahmNatPhys2009} our analysis supports the view that SC and persistent paramagnons may share a common microscopic origin, but need not to strictly couple to each other. The latter statement is consistent with the observation of high-energy paramagnons also in the overdoped regime, where SC is already suppressed.\cite{DeanNatMater2013,JiaNatCommun2014}

We have also noted that the incoherent magnetic excitations are governed predominantly by renormalized single-quasiparticle bandwidth and thus remain sensitive to the magnitude of the on-site Coulomb repulsion. For $U$ much larger than the bare bandwidth, a sharp feature builds up at the threshold of a particle-hole continuum, which is not unambiguously observed in experiments. This circumstance favors the $t$-$J$-$U$ model with intermediate $U$ over the $t$-$J$ model as a starting point for the discussion of the paramagnons in hole-doped high-$T_c$ cuprates. Similar conclusions follow from former studies of equilibrium and single-particle properties.\cite{SpalekPhysRevB2017,ZegrodnikPhysRevB2017,ZegrodnikPhysRevB2017_2,ZegrodnikPhysRevB2018} We have also demonstrated a substantial reduction of charge-mode energy with increasing $U$, which provides the flexibility required to interpret experimental data in real materials. In particular, the above characteristics of collective excitations in the $t$-$J$-$U$ model may allow us to reconcile the observed persistent anti-nodal paramagnons with particularly low-energy acoustic plasmons\cite{NagPhysRevLett2020} in lanthanum cuprates. Finally, the role of $J_\mathrm{eff}$ is expected to be diminished in the regime of extremely large $U$ and small $J$, where ferromagnetic correlations may dominate.\cite{SpalekPhysstatSol1981,EisenbergPhysRevB2002,LiuPhysRevLett2012,BlesioPhysRevB2019} Those aspects should be the subject of a separate study.

A separate future question concerns the influence of the collective excitations on the pairing induced by the correlations.

\section*{Acknowledgments}
 This work  was  supported  by  Grant  OPUS  No.  UMO-2018/29/B/ST3/02646 from Narodowe Centrum Nauki and a grant from the SciMat Priority Research Area under the Strategic Programme Excellence Initiative at the Jagiellonian University.

%%%%%%%%%%%%%%%%%%%%%%%%%%%%%%%%%%%%%%%%%%%%%%%%%%%%%%%%%%%%%%%%%%%%%%%%%%%%%%%%%%%%%%%%%%%%%%%%%%%%%%%%%%%%%%%%%%%%%%%%%%%%%%



\begin{thebibliography}{43}%
\makeatletter
\providecommand \@ifxundefined [1]{%
 \@ifx{#1\undefined}
}%
\providecommand \@ifnum [1]{%
 \ifnum #1\expandafter \@firstoftwo
 \else \expandafter \@secondoftwo
 \fi
}%
\providecommand \@ifx [1]{%
 \ifx #1\expandafter \@firstoftwo
 \else \expandafter \@secondoftwo
 \fi
}%
\providecommand \natexlab [1]{#1}%
\providecommand \enquote  [1]{``#1''}%
\providecommand \bibnamefont  [1]{#1}%
\providecommand \bibfnamefont [1]{#1}%
\providecommand \citenamefont [1]{#1}%
\providecommand \href@noop [0]{\@secondoftwo}%
\providecommand \href [0]{\begingroup \@sanitize@url \@href}%
\providecommand \@href[1]{\@@startlink{#1}\@@href}%
\providecommand \@@href[1]{\endgroup#1\@@endlink}%
\providecommand \@sanitize@url [0]{\catcode `\\12\catcode `\$12\catcode
  `\&12\catcode `\#12\catcode `\^12\catcode `\_12\catcode `\%12\relax}%
\providecommand \@@startlink[1]{}%
\providecommand \@@endlink[0]{}%
\providecommand \url  [0]{\begingroup\@sanitize@url \@url }%
\providecommand \@url [1]{\endgroup\@href {#1}{\urlprefix }}%
\providecommand \urlprefix  [0]{URL }%
\providecommand \Eprint [0]{\href }%
\providecommand \doibase [0]{http://dx.doi.org/}%
\providecommand \selectlanguage [0]{\@gobble}%
\providecommand \bibinfo  [0]{\@secondoftwo}%
\providecommand \bibfield  [0]{\@secondoftwo}%
\providecommand \translation [1]{[#1]}%
\providecommand \BibitemOpen [0]{}%
\providecommand \bibitemStop [0]{}%
\providecommand \bibitemNoStop [0]{.\EOS\space}%
\providecommand \EOS [0]{\spacefactor3000\relax}%
\providecommand \BibitemShut  [1]{\csname bibitem#1\endcsname}%
\let\auto@bib@innerbib\@empty
%</preamble>
\bibitem [{\citenamefont {Spa{\l}ek}\ \emph {et~al.}(2017)\citenamefont
  {Spa{\l}ek}, \citenamefont {Zegrodnik},\ and\ \citenamefont
  {Kaczmarczyk}}]{SpalekPhysRevB2017}%
  \BibitemOpen
  \bibfield  {author} {\bibinfo {author} {\bibfnamefont {J.}~\bibnamefont
  {Spa{\l}ek}}, \bibinfo {author} {\bibfnamefont {M.}~\bibnamefont
  {Zegrodnik}}, \ and\ \bibinfo {author} {\bibfnamefont {J.}~\bibnamefont
  {Kaczmarczyk}},\ }\bibfield  {title} {\enquote {\bibinfo {title} {{Universal
  properties of high-temperature superconductors from real-space pairing:
  $t$-$J$-$U$ model and its quantitative comparison with experiment}},}\ }\href
  {http://dx.doi.org/10.1103/physrevb.95.024506} {\bibfield  {journal}
  {\bibinfo  {journal} {Phys. Rev. B}\ }\textbf {\bibinfo {volume} {95}},\
  \bibinfo {pages} {024506} (\bibinfo {year} {2017})}\BibitemShut {NoStop}%
\bibitem [{\citenamefont {Zegrodnik}\ and\ \citenamefont
  {Spa\l{}ek}(2017{\natexlab{a}})}]{ZegrodnikPhysRevB2017}%
  \BibitemOpen
  \bibfield  {author} {\bibinfo {author} {\bibfnamefont {M.}~\bibnamefont
  {Zegrodnik}}\ and\ \bibinfo {author} {\bibfnamefont {J.}~\bibnamefont
  {Spa\l{}ek}},\ }\bibfield  {title} {\enquote {\bibinfo {title} {{Universal
  properties of high-temperature superconductors from real-space pairing: Role
  of correlated hopping and intersite Coulomb interaction within the
  $t$-$J$-$U$ model}},}\ }\href {https://doi.org/10.1103/PhysRevB.96.054511}
  {\bibfield  {journal} {\bibinfo  {journal} {Phys. Rev. B}\ }\textbf {\bibinfo
  {volume} {96}},\ \bibinfo {pages} {054511} (\bibinfo {year}
  {2017}{\natexlab{a}})}\BibitemShut {NoStop}%
\bibitem [{\citenamefont {Zegrodnik}\ and\ \citenamefont
  {Spa\l{}ek}(2017{\natexlab{b}})}]{ZegrodnikPhysRevB2017_2}%
  \BibitemOpen
  \bibfield  {author} {\bibinfo {author} {\bibfnamefont {M.}~\bibnamefont
  {Zegrodnik}}\ and\ \bibinfo {author} {\bibfnamefont {J.}~\bibnamefont
  {Spa\l{}ek}},\ }\bibfield  {title} {\enquote {\bibinfo {title} {{Effect of
  interlayer processes on the superconducting state within the $t$-$J$-$U$
  model: Full Gutzwiller wave-function solution and relation to experiment}},}\
  }\href {\doibase 10.1103/PhysRevB.95.024507} {\bibfield  {journal} {\bibinfo
  {journal} {Phys. Rev. B}\ }\textbf {\bibinfo {volume} {95}},\ \bibinfo
  {pages} {024507} (\bibinfo {year} {2017}{\natexlab{b}})}\BibitemShut
  {NoStop}%
\bibitem [{\citenamefont {Zegrodnik}\ and\ \citenamefont
  {Spa{\l}ek}(2018)}]{ZegrodnikPhysRevB2018}%
  \BibitemOpen
  \bibfield  {author} {\bibinfo {author} {\bibfnamefont {M.}~\bibnamefont
  {Zegrodnik}}\ and\ \bibinfo {author} {\bibfnamefont {J.}~\bibnamefont
  {Spa{\l}ek}},\ }\bibfield  {title} {\enquote {\bibinfo {title}
  {{Incorporation of charge- and pair-density-wave states into the one-band
  model of $d$-wave superconductivity}},}\ }\href {\doibase
  10.1103/physrevb.98.155144} {\bibfield  {journal} {\bibinfo  {journal} {Phys.
  Rev. B}\ }\textbf {\bibinfo {volume} {98}},\ \bibinfo {pages} {155144}
  (\bibinfo {year} {2018})}\BibitemShut {NoStop}%
\bibitem [{\citenamefont {Dean}\ \emph {et~al.}(2013)\citenamefont {Dean},
  \citenamefont {Dellea}, \citenamefont {Springell}, \citenamefont
  {Yakhou-Harris}, \citenamefont {Kummer}, \citenamefont {Brookes},
  \citenamefont {Liu}, \citenamefont {Sun}, \citenamefont {Strle},
  \citenamefont {Schmitt}, \citenamefont {Braicovich}, \citenamefont
  {Ghiringhelli}, \citenamefont {Bo\v{z}ovi\'c},\ and\ \citenamefont
  {Hill}}]{DeanNatMater2013}%
  \BibitemOpen
  \bibfield  {author} {\bibinfo {author} {\bibfnamefont {M.~P.~M.}\
  \bibnamefont {Dean}}, \bibinfo {author} {\bibfnamefont {G.}~\bibnamefont
  {Dellea}}, \bibinfo {author} {\bibfnamefont {R.~S.}\ \bibnamefont
  {Springell}}, \bibinfo {author} {\bibfnamefont {F.}~\bibnamefont
  {Yakhou-Harris}}, \bibinfo {author} {\bibfnamefont {K.}~\bibnamefont
  {Kummer}}, \bibinfo {author} {\bibfnamefont {N.~B.}\ \bibnamefont {Brookes}},
  \bibinfo {author} {\bibfnamefont {X.}~\bibnamefont {Liu}}, \bibinfo {author}
  {\bibfnamefont {Y-J.}\ \bibnamefont {Sun}}, \bibinfo {author} {\bibfnamefont
  {J.}~\bibnamefont {Strle}}, \bibinfo {author} {\bibfnamefont
  {T.}~\bibnamefont {Schmitt}}, \bibinfo {author} {\bibfnamefont
  {L.}~\bibnamefont {Braicovich}}, \bibinfo {author} {\bibfnamefont
  {G.}~\bibnamefont {Ghiringhelli}}, \bibinfo {author} {\bibfnamefont
  {I.}~\bibnamefont {Bo\v{z}ovi\'c}}, \ and\ \bibinfo {author} {\bibfnamefont
  {J.~P.}\ \bibnamefont {Hill}},\ }\bibfield  {title} {\enquote {\bibinfo
  {title} {{Persistence of magnetic excitations in
  $\mathrm{La_{2-\mathit{x}}Sr_\mathit{x}CuO_4}$ from the undoped insulator to
  the heavily overdoped non-superconducting metal}},}\ }\href
  {http://dx.doi.org/10.1038/nmat3723} {\bibfield  {journal} {\bibinfo
  {journal} {Nat. Mater.}\ }\textbf {\bibinfo {volume} {12}},\ \bibinfo {pages}
  {1019} (\bibinfo {year} {2013})}\BibitemShut {NoStop}%
\bibitem [{\citenamefont {Ishii}\ \emph {et~al.}(2014)\citenamefont {Ishii},
  \citenamefont {Fujita}, \citenamefont {Sasaki}, \citenamefont {Minola},
  \citenamefont {Dellea}, \citenamefont {Mazzoli}, \citenamefont {Kummer},
  \citenamefont {Ghiringhelli}, \citenamefont {Braicovich}, \citenamefont
  {Tohyama}, \citenamefont {Tsutsumi}, \citenamefont {Sato}, \citenamefont
  {Kajimoto}, \citenamefont {Ikeuchi}, \citenamefont {Yamada}, \citenamefont
  {Yoshida}, \citenamefont {Kurooka},\ and\ \citenamefont
  {Mizuki}}]{IshiiNatCommun2014}%
  \BibitemOpen
  \bibfield  {author} {\bibinfo {author} {\bibfnamefont {K.}~\bibnamefont
  {Ishii}}, \bibinfo {author} {\bibfnamefont {M.}~\bibnamefont {Fujita}},
  \bibinfo {author} {\bibfnamefont {T.}~\bibnamefont {Sasaki}}, \bibinfo
  {author} {\bibfnamefont {M.}~\bibnamefont {Minola}}, \bibinfo {author}
  {\bibfnamefont {G.}~\bibnamefont {Dellea}}, \bibinfo {author} {\bibfnamefont
  {C.}~\bibnamefont {Mazzoli}}, \bibinfo {author} {\bibfnamefont
  {K.}~\bibnamefont {Kummer}}, \bibinfo {author} {\bibfnamefont
  {G.}~\bibnamefont {Ghiringhelli}}, \bibinfo {author} {\bibfnamefont
  {L.}~\bibnamefont {Braicovich}}, \bibinfo {author} {\bibfnamefont
  {T.}~\bibnamefont {Tohyama}}, \bibinfo {author} {\bibfnamefont
  {K.}~\bibnamefont {Tsutsumi}}, \bibinfo {author} {\bibfnamefont
  {K.}~\bibnamefont {Sato}}, \bibinfo {author} {\bibfnamefont {R.}~\bibnamefont
  {Kajimoto}}, \bibinfo {author} {\bibfnamefont {K.}~\bibnamefont {Ikeuchi}},
  \bibinfo {author} {\bibfnamefont {K.}~\bibnamefont {Yamada}}, \bibinfo
  {author} {\bibfnamefont {M.}~\bibnamefont {Yoshida}}, \bibinfo {author}
  {\bibfnamefont {M.}~\bibnamefont {Kurooka}}, \ and\ \bibinfo {author}
  {\bibfnamefont {J.}~\bibnamefont {Mizuki}},\ }\bibfield  {title} {\enquote
  {\bibinfo {title} {{High-energy spin and charge excitations in electron-doped
  copper oxide superconductors}},}\ }\href
  {http://dx.doi.org/10.1038/ncomms4714} {\bibfield  {journal} {\bibinfo
  {journal} {Nat. Commun.}\ }\textbf {\bibinfo {volume} {5}},\ \bibinfo {pages}
  {3714} (\bibinfo {year} {2014})}\BibitemShut {NoStop}%
\bibitem [{\citenamefont {Lee}\ \emph {et~al.}(2014)\citenamefont {Lee},
  \citenamefont {Lee}, \citenamefont {Nowadnick}, \citenamefont {Gerber},
  \citenamefont {Tabi\'{s}}, \citenamefont {Huang}, \citenamefont {Strocov},
  \citenamefont {Motoyama}, \citenamefont {Yu}, \citenamefont {Moritz},
  \citenamefont {Huang}, \citenamefont {Wang}, \citenamefont {Huang},
  \citenamefont {Wu}, \citenamefont {Chen}, \citenamefont {Huang},
  \citenamefont {Greven}, \citenamefont {Schmitt}, \citenamefont {Shen},\ and\
  \citenamefont {Devereaux}}]{LeeNatPhys2014}%
  \BibitemOpen
  \bibfield  {author} {\bibinfo {author} {\bibfnamefont {W.~S.}\ \bibnamefont
  {Lee}}, \bibinfo {author} {\bibfnamefont {J.~J.}\ \bibnamefont {Lee}},
  \bibinfo {author} {\bibfnamefont {E.~A.}\ \bibnamefont {Nowadnick}}, \bibinfo
  {author} {\bibfnamefont {S.}~\bibnamefont {Gerber}}, \bibinfo {author}
  {\bibfnamefont {W.}~\bibnamefont {Tabi\'{s}}}, \bibinfo {author}
  {\bibfnamefont {S.~W.}\ \bibnamefont {Huang}}, \bibinfo {author}
  {\bibfnamefont {V.~N.}\ \bibnamefont {Strocov}}, \bibinfo {author}
  {\bibfnamefont {E.~M.}\ \bibnamefont {Motoyama}}, \bibinfo {author}
  {\bibfnamefont {G.}~\bibnamefont {Yu}}, \bibinfo {author} {\bibfnamefont
  {B.}~\bibnamefont {Moritz}}, \bibinfo {author} {\bibfnamefont {H.~Y.}\
  \bibnamefont {Huang}}, \bibinfo {author} {\bibfnamefont {R.~P.}\ \bibnamefont
  {Wang}}, \bibinfo {author} {\bibfnamefont {Y.~B.}\ \bibnamefont {Huang}},
  \bibinfo {author} {\bibfnamefont {W.~B.}\ \bibnamefont {Wu}}, \bibinfo
  {author} {\bibfnamefont {C.~T.}\ \bibnamefont {Chen}}, \bibinfo {author}
  {\bibfnamefont {D.~J.}\ \bibnamefont {Huang}}, \bibinfo {author}
  {\bibfnamefont {M.}~\bibnamefont {Greven}}, \bibinfo {author} {\bibfnamefont
  {T.}~\bibnamefont {Schmitt}}, \bibinfo {author} {\bibfnamefont {Z.~X.}\
  \bibnamefont {Shen}}, \ and\ \bibinfo {author} {\bibfnamefont {T.~P.}\
  \bibnamefont {Devereaux}},\ }\bibfield  {title} {\enquote {\bibinfo {title}
  {{Asymmetry of collective excitations in electron- and hole-doped cuprate
  superconductors}},}\ }\href {http://dx.doi.org/10.1038/nphys3117} {\bibfield
  {journal} {\bibinfo  {journal} {Nat. Phys.}\ }\textbf {\bibinfo {volume}
  {10}},\ \bibinfo {pages} {883} (\bibinfo {year} {2014})}\BibitemShut
  {NoStop}%
\bibitem [{\citenamefont {Guarise}\ \emph {et~al.}(2014)\citenamefont
  {Guarise}, \citenamefont {Piazza}, \citenamefont {Berger}, \citenamefont
  {Giannini}, \citenamefont {Schmitt}, \citenamefont {R{\o}nnow}, \citenamefont
  {Sawatzky}, \citenamefont {van~den Brink}, \citenamefont {Altenfeld},
  \citenamefont {Eremin},\ and\ \citenamefont {Grioni}}]{GuariseNatCommun2014}%
  \BibitemOpen
  \bibfield  {author} {\bibinfo {author} {\bibfnamefont {M.}~\bibnamefont
  {Guarise}}, \bibinfo {author} {\bibfnamefont {B.~Dalla}\ \bibnamefont
  {Piazza}}, \bibinfo {author} {\bibfnamefont {H.}~\bibnamefont {Berger}},
  \bibinfo {author} {\bibfnamefont {E.}~\bibnamefont {Giannini}}, \bibinfo
  {author} {\bibfnamefont {T.}~\bibnamefont {Schmitt}}, \bibinfo {author}
  {\bibfnamefont {H.~M.}\ \bibnamefont {R{\o}nnow}}, \bibinfo {author}
  {\bibfnamefont {G.~A.}\ \bibnamefont {Sawatzky}}, \bibinfo {author}
  {\bibfnamefont {J.}~\bibnamefont {van~den Brink}}, \bibinfo {author}
  {\bibfnamefont {D.}~\bibnamefont {Altenfeld}}, \bibinfo {author}
  {\bibfnamefont {I.}~\bibnamefont {Eremin}}, \ and\ \bibinfo {author}
  {\bibfnamefont {M.}~\bibnamefont {Grioni}},\ }\bibfield  {title} {\enquote
  {\bibinfo {title} {{Anisotropic softening of magnetic excitations along the
  nodal direction in superconducting cuprates}},}\ }\href
  {http://dx.doi.org/10.1038/ncomms6760} {\bibfield  {journal} {\bibinfo
  {journal} {Nat. Commun.}\ }\textbf {\bibinfo {volume} {5}},\ \bibinfo {pages}
  {5760} (\bibinfo {year} {2014})}\BibitemShut {NoStop}%
\bibitem [{\citenamefont {Wakimoto}\ \emph {et~al.}(2015)\citenamefont
  {Wakimoto}, \citenamefont {Ishii}, \citenamefont {Kimura}, \citenamefont
  {Fujita}, \citenamefont {Dellea}, \citenamefont {Kummer}, \citenamefont
  {Braicovich}, \citenamefont {Ghiringhelli}, \citenamefont {Debeer-Schmitt},\
  and\ \citenamefont {Granroth}}]{WakimotoPhysRevB2015}%
  \BibitemOpen
  \bibfield  {author} {\bibinfo {author} {\bibfnamefont {S.}~\bibnamefont
  {Wakimoto}}, \bibinfo {author} {\bibfnamefont {K.}~\bibnamefont {Ishii}},
  \bibinfo {author} {\bibfnamefont {H.}~\bibnamefont {Kimura}}, \bibinfo
  {author} {\bibfnamefont {M.}~\bibnamefont {Fujita}}, \bibinfo {author}
  {\bibfnamefont {G.}~\bibnamefont {Dellea}}, \bibinfo {author} {\bibfnamefont
  {K.}~\bibnamefont {Kummer}}, \bibinfo {author} {\bibfnamefont
  {L.}~\bibnamefont {Braicovich}}, \bibinfo {author} {\bibfnamefont
  {G.}~\bibnamefont {Ghiringhelli}}, \bibinfo {author} {\bibfnamefont {L.~M.}\
  \bibnamefont {Debeer-Schmitt}}, \ and\ \bibinfo {author} {\bibfnamefont
  {G.~E.}\ \bibnamefont {Granroth}},\ }\bibfield  {title} {\enquote {\bibinfo
  {title} {{High-energy magnetic excitations in overdoped
  $\mathrm{La_{2-\mathit{x}}Sr_\mathit{x}CuO_4}$ studied by neutron and
  resonant inelastic $x$-ray scattering}},}\ }\href
  {http://dx.doi.org/10.1103/physrevb.91.184513} {\bibfield  {journal}
  {\bibinfo  {journal} {Phys. Rev. B}\ }\textbf {\bibinfo {volume} {91}}
  (\bibinfo {year} {2015})}\BibitemShut {NoStop}%
\bibitem [{\citenamefont {Minola}\ \emph {et~al.}(2017)\citenamefont {Minola},
  \citenamefont {Lu}, \citenamefont {Peng}, \citenamefont {Dellea},
  \citenamefont {Gretarsson}, \citenamefont {Haverkort}, \citenamefont {Ding},
  \citenamefont {Sun}, \citenamefont {Zhou}, \citenamefont {Peets},
  \citenamefont {Chauviere}, \citenamefont {Dosanjh}, \citenamefont {Bonn},
  \citenamefont {Liang}, \citenamefont {Damascelli}, \citenamefont {Dantz},
  \citenamefont {Lu}, \citenamefont {Schmitt}, \citenamefont {Braicovich},
  \citenamefont {Ghiringhelli}, \citenamefont {Keimer},\ and\ \citenamefont
  {Le~Tacon}}]{MinolaPhysRevLett2017}%
  \BibitemOpen
  \bibfield  {author} {\bibinfo {author} {\bibfnamefont {M.}~\bibnamefont
  {Minola}}, \bibinfo {author} {\bibfnamefont {Y.}~\bibnamefont {Lu}}, \bibinfo
  {author} {\bibfnamefont {Y.~Y.}\ \bibnamefont {Peng}}, \bibinfo {author}
  {\bibfnamefont {G.}~\bibnamefont {Dellea}}, \bibinfo {author} {\bibfnamefont
  {H.}~\bibnamefont {Gretarsson}}, \bibinfo {author} {\bibfnamefont {M.~W.}\
  \bibnamefont {Haverkort}}, \bibinfo {author} {\bibfnamefont {Y.}~\bibnamefont
  {Ding}}, \bibinfo {author} {\bibfnamefont {X.}~\bibnamefont {Sun}}, \bibinfo
  {author} {\bibfnamefont {X.~J.}\ \bibnamefont {Zhou}}, \bibinfo {author}
  {\bibfnamefont {D.~C.}\ \bibnamefont {Peets}}, \bibinfo {author}
  {\bibfnamefont {L.}~\bibnamefont {Chauviere}}, \bibinfo {author}
  {\bibfnamefont {P.}~\bibnamefont {Dosanjh}}, \bibinfo {author} {\bibfnamefont
  {D.~A.}\ \bibnamefont {Bonn}}, \bibinfo {author} {\bibfnamefont
  {R.}~\bibnamefont {Liang}}, \bibinfo {author} {\bibfnamefont
  {A.}~\bibnamefont {Damascelli}}, \bibinfo {author} {\bibfnamefont
  {M.}~\bibnamefont {Dantz}}, \bibinfo {author} {\bibfnamefont
  {X.}~\bibnamefont {Lu}}, \bibinfo {author} {\bibfnamefont {T.}~\bibnamefont
  {Schmitt}}, \bibinfo {author} {\bibfnamefont {L.}~\bibnamefont {Braicovich}},
  \bibinfo {author} {\bibfnamefont {G.}~\bibnamefont {Ghiringhelli}}, \bibinfo
  {author} {\bibfnamefont {B.}~\bibnamefont {Keimer}}, \ and\ \bibinfo {author}
  {\bibfnamefont {M.}~\bibnamefont {Le~Tacon}},\ }\bibfield  {title} {\enquote
  {\bibinfo {title} {{Crossover from Collective to Incoherent Spin Excitations
  in Superconducting Cuprates Probed by Detuned Resonant Inelastic $X$-Ray
  Scattering}},}\ }\href {http://dx.doi.org/10.1103/physrevlett.119.097001}
  {\bibfield  {journal} {\bibinfo  {journal} {Phys. Rev. Lett.}\ }\textbf
  {\bibinfo {volume} {119}},\ \bibinfo {pages} {245133} (\bibinfo {year}
  {2017})}\BibitemShut {NoStop}%
\bibitem [{\citenamefont {Ivashko}\ \emph {et~al.}(2017)\citenamefont
  {Ivashko}, \citenamefont {Shaik}, \citenamefont {Lu}, \citenamefont
  {Fatuzzo}, \citenamefont {Dantz}, \citenamefont {Freeman}, \citenamefont
  {McNally}, \citenamefont {Destraz}, \citenamefont {Christensen},
  \citenamefont {Kurosawa}, \citenamefont {Momono}, \citenamefont {Oda},
  \citenamefont {Matt}, \citenamefont {Monney}, \citenamefont {R\o{}nnow},
  \citenamefont {Schmitt},\ and\ \citenamefont {Chang}}]{IvashkoPhysRevB2017}%
  \BibitemOpen
  \bibfield  {author} {\bibinfo {author} {\bibfnamefont {O.}~\bibnamefont
  {Ivashko}}, \bibinfo {author} {\bibfnamefont {N.~E.}\ \bibnamefont {Shaik}},
  \bibinfo {author} {\bibfnamefont {X.}~\bibnamefont {Lu}}, \bibinfo {author}
  {\bibfnamefont {C.~G.}\ \bibnamefont {Fatuzzo}}, \bibinfo {author}
  {\bibfnamefont {M.}~\bibnamefont {Dantz}}, \bibinfo {author} {\bibfnamefont
  {P.~G.}\ \bibnamefont {Freeman}}, \bibinfo {author} {\bibfnamefont {D.~E.}\
  \bibnamefont {McNally}}, \bibinfo {author} {\bibfnamefont {D.}~\bibnamefont
  {Destraz}}, \bibinfo {author} {\bibfnamefont {N.~B.}\ \bibnamefont
  {Christensen}}, \bibinfo {author} {\bibfnamefont {T.}~\bibnamefont
  {Kurosawa}}, \bibinfo {author} {\bibfnamefont {N.}~\bibnamefont {Momono}},
  \bibinfo {author} {\bibfnamefont {M.}~\bibnamefont {Oda}}, \bibinfo {author}
  {\bibfnamefont {C.~E.}\ \bibnamefont {Matt}}, \bibinfo {author}
  {\bibfnamefont {C.}~\bibnamefont {Monney}}, \bibinfo {author} {\bibfnamefont
  {H.~M.}\ \bibnamefont {R\o{}nnow}}, \bibinfo {author} {\bibfnamefont
  {T.}~\bibnamefont {Schmitt}}, \ and\ \bibinfo {author} {\bibfnamefont
  {J.}~\bibnamefont {Chang}},\ }\bibfield  {title} {\enquote {\bibinfo {title}
  {{Damped spin excitations in a doped cuprate superconductor with orbital
  hybridization}},}\ }\href {http://dx.doi.org/10.1103/physrevb.95.214508}
  {\bibfield  {journal} {\bibinfo  {journal} {Phys. Rev. B}\ }\textbf {\bibinfo
  {volume} {95}} (\bibinfo {year} {2017})}\BibitemShut {NoStop}%
\bibitem [{\citenamefont {Meyers}\ \emph {et~al.}(2017)\citenamefont {Meyers},
  \citenamefont {Miao}, \citenamefont {Walters}, \citenamefont {Bisogni},
  \citenamefont {Springell}, \citenamefont {d'Astuto}, \citenamefont {Dantz},
  \citenamefont {Pelliciari}, \citenamefont {Huang}, \citenamefont {Okamoto},
  \citenamefont {Huang}, \citenamefont {Hill}, \citenamefont {He},
  \citenamefont {Bo\ifmmode \check{z}\else \v{z}\fi{}ovi\ifmmode~\acute{c}\else
  \'{c}\fi{}}, \citenamefont {Schmitt},\ and\ \citenamefont
  {Dean}}]{MeyersPhysRevB2017}%
  \BibitemOpen
  \bibfield  {author} {\bibinfo {author} {\bibfnamefont {D.}~\bibnamefont
  {Meyers}}, \bibinfo {author} {\bibfnamefont {H.}~\bibnamefont {Miao}},
  \bibinfo {author} {\bibfnamefont {A.~C.}\ \bibnamefont {Walters}}, \bibinfo
  {author} {\bibfnamefont {V.}~\bibnamefont {Bisogni}}, \bibinfo {author}
  {\bibfnamefont {R.~S.}\ \bibnamefont {Springell}}, \bibinfo {author}
  {\bibfnamefont {M.}~\bibnamefont {d'Astuto}}, \bibinfo {author}
  {\bibfnamefont {M.}~\bibnamefont {Dantz}}, \bibinfo {author} {\bibfnamefont
  {J.}~\bibnamefont {Pelliciari}}, \bibinfo {author} {\bibfnamefont {H.~Y.}\
  \bibnamefont {Huang}}, \bibinfo {author} {\bibfnamefont {J.}~\bibnamefont
  {Okamoto}}, \bibinfo {author} {\bibfnamefont {D.~J.}\ \bibnamefont {Huang}},
  \bibinfo {author} {\bibfnamefont {J.~P.}\ \bibnamefont {Hill}}, \bibinfo
  {author} {\bibfnamefont {X.}~\bibnamefont {He}}, \bibinfo {author}
  {\bibfnamefont {I.}~\bibnamefont {Bo\ifmmode \check{z}\else
  \v{z}\fi{}ovi\ifmmode~\acute{c}\else \'{c}\fi{}}}, \bibinfo {author}
  {\bibfnamefont {T.}~\bibnamefont {Schmitt}}, \ and\ \bibinfo {author}
  {\bibfnamefont {M.~P.~M.}\ \bibnamefont {Dean}},\ }\bibfield  {title}
  {\enquote {\bibinfo {title} {{Doping dependence of the magnetic excitations
  in $\mathrm{La_{2-\mathit{x}}Sr_\mathit{x}CuO_4}$}},}\ }\href
  {http://dx.doi.org/10.1103/PhysRevB.95.075139} {\bibfield  {journal}
  {\bibinfo  {journal} {Phys. Rev. B}\ }\textbf {\bibinfo {volume} {95}},\
  \bibinfo {pages} {075139} (\bibinfo {year} {2017})}\BibitemShut {NoStop}%
\bibitem [{\citenamefont {Chaix}\ \emph {et~al.}(2018)\citenamefont {Chaix},
  \citenamefont {Huang}, \citenamefont {Gerber}, \citenamefont {Lu},
  \citenamefont {Jia}, \citenamefont {Huang}, \citenamefont {McNally},
  \citenamefont {Wang}, \citenamefont {Vernay}, \citenamefont {Keren},
  \citenamefont {Shi}, \citenamefont {Moritz}, \citenamefont {Shen},
  \citenamefont {Schmitt}, \citenamefont {Devereaux},\ and\ \citenamefont
  {Lee}}]{ChaixPhysRevB2018}%
  \BibitemOpen
  \bibfield  {author} {\bibinfo {author} {\bibfnamefont {L.}~\bibnamefont
  {Chaix}}, \bibinfo {author} {\bibfnamefont {E.~W.}\ \bibnamefont {Huang}},
  \bibinfo {author} {\bibfnamefont {S.}~\bibnamefont {Gerber}}, \bibinfo
  {author} {\bibfnamefont {X.}~\bibnamefont {Lu}}, \bibinfo {author}
  {\bibfnamefont {C.}~\bibnamefont {Jia}}, \bibinfo {author} {\bibfnamefont
  {Y.}~\bibnamefont {Huang}}, \bibinfo {author} {\bibfnamefont {D.~E.}\
  \bibnamefont {McNally}}, \bibinfo {author} {\bibfnamefont {Y.}~\bibnamefont
  {Wang}}, \bibinfo {author} {\bibfnamefont {F.~H.}\ \bibnamefont {Vernay}},
  \bibinfo {author} {\bibfnamefont {A.}~\bibnamefont {Keren}}, \bibinfo
  {author} {\bibfnamefont {M.}~\bibnamefont {Shi}}, \bibinfo {author}
  {\bibfnamefont {B.}~\bibnamefont {Moritz}}, \bibinfo {author} {\bibfnamefont
  {Z.-X.}\ \bibnamefont {Shen}}, \bibinfo {author} {\bibfnamefont
  {T.}~\bibnamefont {Schmitt}}, \bibinfo {author} {\bibfnamefont {T.~P.}\
  \bibnamefont {Devereaux}}, \ and\ \bibinfo {author} {\bibfnamefont {W.-S.}\
  \bibnamefont {Lee}},\ }\bibfield  {title} {\enquote {\bibinfo {title}
  {{Resonant inelastic $x$-ray scattering studies of magnons band bimagnons in
  the lightly doped cuprate $\mathrm{La_{2-\mathit{x}}Sr_\mathit{x}CuO_4}$}},}\
  }\href {http://dx.doi.org/10.1103/physrevb.97.155144} {\bibfield  {journal}
  {\bibinfo  {journal} {Phys. Rev. B}\ }\textbf {\bibinfo {volume} {97}},\
  \bibinfo {pages} {155144} (\bibinfo {year} {2018})}\BibitemShut {NoStop}%
\bibitem [{\citenamefont {Robarts}\ \emph {et~al.}(2019)\citenamefont
  {Robarts}, \citenamefont {Barth\'elemy}, \citenamefont {Kummer},
  \citenamefont {Garc\'{\i}a-Fern\'andez}, \citenamefont {Li}, \citenamefont
  {Nag}, \citenamefont {Walters}, \citenamefont {Zhou},\ and\ \citenamefont
  {Hayden}}]{Robarts_arXiV_2019}%
  \BibitemOpen
  \bibfield  {author} {\bibinfo {author} {\bibfnamefont {H.~C.}\ \bibnamefont
  {Robarts}}, \bibinfo {author} {\bibfnamefont {M.}~\bibnamefont
  {Barth\'elemy}}, \bibinfo {author} {\bibfnamefont {K.}~\bibnamefont
  {Kummer}}, \bibinfo {author} {\bibfnamefont {M.}~\bibnamefont
  {Garc\'{\i}a-Fern\'andez}}, \bibinfo {author} {\bibfnamefont
  {J.}~\bibnamefont {Li}}, \bibinfo {author} {\bibfnamefont {A.}~\bibnamefont
  {Nag}}, \bibinfo {author} {\bibfnamefont {A.~C.}\ \bibnamefont {Walters}},
  \bibinfo {author} {\bibfnamefont {K.~J.}\ \bibnamefont {Zhou}}, \ and\
  \bibinfo {author} {\bibfnamefont {S.~M.}\ \bibnamefont {Hayden}},\ }\bibfield
   {title} {\enquote {\bibinfo {title} {{Anisotropic damping and wave vector
  dependent susceptibility of the spin fluctuations in
  ${\mathrm{La}}_{2\ensuremath{-}x}{\mathrm{Sr}}_{x}{\mathrm{CuO}}_{4}$ studied
  by resonant inelastic x-ray scattering}},}\ }\href {\doibase
  10.1103/PhysRevB.100.214510} {\bibfield  {journal} {\bibinfo  {journal}
  {Phys. Rev. B}\ }\textbf {\bibinfo {volume} {100}},\ \bibinfo {pages}
  {214510} (\bibinfo {year} {2019})}\BibitemShut {NoStop}%
\bibitem [{\citenamefont {Zhou}\ \emph {et~al.}(2013)\citenamefont {Zhou},
  \citenamefont {Huang}, \citenamefont {Monney}, \citenamefont {Dai},
  \citenamefont {Strocov}, \citenamefont {Wang}, \citenamefont {Chen},
  \citenamefont {Zhang}, \citenamefont {Dai}, \citenamefont {Patthey},
  \citenamefont {van~den Brink}, \citenamefont {Ding},\ and\ \citenamefont
  {Schmitt}}]{ZhouNatCommun2013}%
  \BibitemOpen
  \bibfield  {author} {\bibinfo {author} {\bibfnamefont {K.-J.}\ \bibnamefont
  {Zhou}}, \bibinfo {author} {\bibfnamefont {Y.-B.}\ \bibnamefont {Huang}},
  \bibinfo {author} {\bibfnamefont {C.}~\bibnamefont {Monney}}, \bibinfo
  {author} {\bibfnamefont {X.}~\bibnamefont {Dai}}, \bibinfo {author}
  {\bibfnamefont {V.~N.}\ \bibnamefont {Strocov}}, \bibinfo {author}
  {\bibfnamefont {N.-L.}\ \bibnamefont {Wang}}, \bibinfo {author}
  {\bibfnamefont {Z.-G.}\ \bibnamefont {Chen}}, \bibinfo {author}
  {\bibfnamefont {C.}~\bibnamefont {Zhang}}, \bibinfo {author} {\bibfnamefont
  {P.}~\bibnamefont {Dai}}, \bibinfo {author} {\bibfnamefont {L.}~\bibnamefont
  {Patthey}}, \bibinfo {author} {\bibfnamefont {J.}~\bibnamefont {van~den
  Brink}}, \bibinfo {author} {\bibfnamefont {H.}~\bibnamefont {Ding}}, \ and\
  \bibinfo {author} {\bibfnamefont {T.}~\bibnamefont {Schmitt}},\ }\bibfield
  {title} {\enquote {\bibinfo {title} {{Persistent high-energy spin excitations
  in iron-pnictide superconductors}},}\ }\href
  {http://dx.doi.org/10.1038/ncomms2428} {\bibfield  {journal} {\bibinfo
  {journal} {Nat. Commun.}\ }\textbf {\bibinfo {volume} {4}},\ \bibinfo {pages}
  {1470} (\bibinfo {year} {2013})}\BibitemShut {NoStop}%
\bibitem [{\citenamefont {Gretarsson}\ \emph {et~al.}(2016)\citenamefont
  {Gretarsson}, \citenamefont {Sung}, \citenamefont {Porras}, \citenamefont
  {Bertinshaw}, \citenamefont {Dietl}, \citenamefont {Bruin}, \citenamefont
  {Bangura}, \citenamefont {Kim}, \citenamefont {Dinnebier}, \citenamefont
  {Kim}, \citenamefont {Al-Zein}, \citenamefont {Moretti~Sala}, \citenamefont
  {Krisch}, \citenamefont {Le~Tacon}, \citenamefont {Keimer},\ and\
  \citenamefont {Kim}}]{GretarssonPhysRevLett2016}%
  \BibitemOpen
  \bibfield  {author} {\bibinfo {author} {\bibfnamefont {H.}~\bibnamefont
  {Gretarsson}}, \bibinfo {author} {\bibfnamefont {N.~H.}\ \bibnamefont
  {Sung}}, \bibinfo {author} {\bibfnamefont {J.}~\bibnamefont {Porras}},
  \bibinfo {author} {\bibfnamefont {J.}~\bibnamefont {Bertinshaw}}, \bibinfo
  {author} {\bibfnamefont {C.}~\bibnamefont {Dietl}}, \bibinfo {author}
  {\bibfnamefont {Jan A.~N.}\ \bibnamefont {Bruin}}, \bibinfo {author}
  {\bibfnamefont {A.~F.}\ \bibnamefont {Bangura}}, \bibinfo {author}
  {\bibfnamefont {Y.~K.}\ \bibnamefont {Kim}}, \bibinfo {author} {\bibfnamefont
  {R.}~\bibnamefont {Dinnebier}}, \bibinfo {author} {\bibfnamefont {Jungho}\
  \bibnamefont {Kim}}, \bibinfo {author} {\bibfnamefont {A.}~\bibnamefont
  {Al-Zein}}, \bibinfo {author} {\bibfnamefont {M.}~\bibnamefont
  {Moretti~Sala}}, \bibinfo {author} {\bibfnamefont {M.}~\bibnamefont
  {Krisch}}, \bibinfo {author} {\bibfnamefont {M.}~\bibnamefont {Le~Tacon}},
  \bibinfo {author} {\bibfnamefont {B.}~\bibnamefont {Keimer}}, \ and\ \bibinfo
  {author} {\bibfnamefont {B.~J.}\ \bibnamefont {Kim}},\ }\bibfield  {title}
  {\enquote {\bibinfo {title} {{Persistent Paramagnons Deep in the Metallic
  Phase of $\mathrm{Sr_{2-\mathit{x}}La_\mathit{x}IrO_4}$}},}\ }\href
  {http://dx.doi.org/10.1103/PhysRevLett.117.107001} {\bibfield  {journal}
  {\bibinfo  {journal} {Phys. Rev. Lett.}\ }\textbf {\bibinfo {volume} {117}},\
  \bibinfo {pages} {107001} (\bibinfo {year} {2016})}\BibitemShut {NoStop}%
\bibitem [{\citenamefont {Fumagalli}\ \emph {et~al.}(2019)\citenamefont
  {Fumagalli}, \citenamefont {Braicovich}, \citenamefont {Minola},
  \citenamefont {Peng}, \citenamefont {Kummer}, \citenamefont {Betto},
  \citenamefont {Rossi}, \citenamefont {Lefran\ifmmode~\mbox{\c{c}}\else
  \c{c}\fi{}ois}, \citenamefont {Morawe}, \citenamefont {Salluzzo},
  \citenamefont {Suzuki}, \citenamefont {Yakhou}, \citenamefont {Le~Tacon},
  \citenamefont {Keimer}, \citenamefont {Brookes}, \citenamefont {Sala},\ and\
  \citenamefont {Ghiringhelli}}]{FumagalliPhysRevB2019}%
  \BibitemOpen
  \bibfield  {author} {\bibinfo {author} {\bibfnamefont {R.}~\bibnamefont
  {Fumagalli}}, \bibinfo {author} {\bibfnamefont {L.}~\bibnamefont
  {Braicovich}}, \bibinfo {author} {\bibfnamefont {M.}~\bibnamefont {Minola}},
  \bibinfo {author} {\bibfnamefont {Y.~Y.}\ \bibnamefont {Peng}}, \bibinfo
  {author} {\bibfnamefont {K.}~\bibnamefont {Kummer}}, \bibinfo {author}
  {\bibfnamefont {D.}~\bibnamefont {Betto}}, \bibinfo {author} {\bibfnamefont
  {M.}~\bibnamefont {Rossi}}, \bibinfo {author} {\bibfnamefont
  {E.}~\bibnamefont {Lefran\ifmmode~\mbox{\c{c}}\else \c{c}\fi{}ois}}, \bibinfo
  {author} {\bibfnamefont {C.}~\bibnamefont {Morawe}}, \bibinfo {author}
  {\bibfnamefont {M.}~\bibnamefont {Salluzzo}}, \bibinfo {author}
  {\bibfnamefont {H.}~\bibnamefont {Suzuki}}, \bibinfo {author} {\bibfnamefont
  {F.}~\bibnamefont {Yakhou}}, \bibinfo {author} {\bibfnamefont
  {M.}~\bibnamefont {Le~Tacon}}, \bibinfo {author} {\bibfnamefont
  {B.}~\bibnamefont {Keimer}}, \bibinfo {author} {\bibfnamefont {N.~B.}\
  \bibnamefont {Brookes}}, \bibinfo {author} {\bibfnamefont {M.~Moretti}\
  \bibnamefont {Sala}}, \ and\ \bibinfo {author} {\bibfnamefont
  {G.}~\bibnamefont {Ghiringhelli}},\ }\bibfield  {title} {\enquote {\bibinfo
  {title} {{Polarization-resolved Cu $L_3$-edge resonant inelastic $x$-ray
  scattering of orbital and spin excitations in
  $\mathrm{NdBa_2Cu_3O_{7-\delta}}$}},}\ }\href {\doibase
  10.1103/physrevb.99.134517} {\bibfield  {journal} {\bibinfo  {journal} {Phys.
  Rev. B}\ }\textbf {\bibinfo {volume} {99}},\ \bibinfo {pages} {134517}
  (\bibinfo {year} {2019})}\BibitemShut {NoStop}%
\bibitem [{\citenamefont {Le~Tacon}\ \emph {et~al.}(2011)\citenamefont
  {Le~Tacon}, \citenamefont {Ghiringhelli}, \citenamefont {Chaloupka},
  \citenamefont {Sala}, \citenamefont {Hinkov}, \citenamefont {Haverkort},
  \citenamefont {Minola}, \citenamefont {Bakr}, \citenamefont {Zhou},
  \citenamefont {Blanco-Canosa}, \citenamefont {Monney}, \citenamefont {Song},
  \citenamefont {Sun}, \citenamefont {Lin}, \citenamefont {De~Luca},
  \citenamefont {Salluzzo}, \citenamefont {Khaliullin}, \citenamefont
  {Schmitt}, \citenamefont {Braicovich},\ and\ \citenamefont
  {Keimer}}]{LeTaconNatPhys2011}%
  \BibitemOpen
  \bibfield  {author} {\bibinfo {author} {\bibfnamefont {M.}~\bibnamefont
  {Le~Tacon}}, \bibinfo {author} {\bibfnamefont {G.}~\bibnamefont
  {Ghiringhelli}}, \bibinfo {author} {\bibfnamefont {J.}~\bibnamefont
  {Chaloupka}}, \bibinfo {author} {\bibfnamefont {M.~Moretti}\ \bibnamefont
  {Sala}}, \bibinfo {author} {\bibfnamefont {V.}~\bibnamefont {Hinkov}},
  \bibinfo {author} {\bibfnamefont {M.~W.}\ \bibnamefont {Haverkort}}, \bibinfo
  {author} {\bibfnamefont {M.}~\bibnamefont {Minola}}, \bibinfo {author}
  {\bibfnamefont {M.}~\bibnamefont {Bakr}}, \bibinfo {author} {\bibfnamefont
  {K.~J.}\ \bibnamefont {Zhou}}, \bibinfo {author} {\bibfnamefont
  {S.}~\bibnamefont {Blanco-Canosa}}, \bibinfo {author} {\bibfnamefont
  {C.}~\bibnamefont {Monney}}, \bibinfo {author} {\bibfnamefont {Y.~T.}\
  \bibnamefont {Song}}, \bibinfo {author} {\bibfnamefont {G.~L.}\ \bibnamefont
  {Sun}}, \bibinfo {author} {\bibfnamefont {C.~T.}\ \bibnamefont {Lin}},
  \bibinfo {author} {\bibfnamefont {G.~M.}\ \bibnamefont {De~Luca}}, \bibinfo
  {author} {\bibfnamefont {M.}~\bibnamefont {Salluzzo}}, \bibinfo {author}
  {\bibfnamefont {G.}~\bibnamefont {Khaliullin}}, \bibinfo {author}
  {\bibfnamefont {T.}~\bibnamefont {Schmitt}}, \bibinfo {author} {\bibfnamefont
  {L.}~\bibnamefont {Braicovich}}, \ and\ \bibinfo {author} {\bibfnamefont
  {B.}~\bibnamefont {Keimer}},\ }\bibfield  {title} {\enquote {\bibinfo {title}
  {{Intense paramagnon excitations in a large family of high-temperature
  superconductors}},}\ }\href {http://dx.doi.org/10.1038/nphys2041} {\bibfield
  {journal} {\bibinfo  {journal} {Nat. Phys.}\ }\textbf {\bibinfo {volume}
  {7}},\ \bibinfo {pages} {725} (\bibinfo {year} {2011})}\BibitemShut {NoStop}%
\bibitem [{\citenamefont {Jia}\ \emph {et~al.}(2014)\citenamefont {Jia},
  \citenamefont {Nowadnick}, \citenamefont {Wohlfeld}, \citenamefont {Kung},
  \citenamefont {Chen}, \citenamefont {Johnston}, \citenamefont {Tohyama},
  \citenamefont {Moritz},\ and\ \citenamefont {Devereaux}}]{JiaNatCommun2014}%
  \BibitemOpen
  \bibfield  {author} {\bibinfo {author} {\bibfnamefont {C.~J.}\ \bibnamefont
  {Jia}}, \bibinfo {author} {\bibfnamefont {E.~A.}\ \bibnamefont {Nowadnick}},
  \bibinfo {author} {\bibfnamefont {K.}~\bibnamefont {Wohlfeld}}, \bibinfo
  {author} {\bibfnamefont {Y.~F.}\ \bibnamefont {Kung}}, \bibinfo {author}
  {\bibfnamefont {C.-C.}\ \bibnamefont {Chen}}, \bibinfo {author}
  {\bibfnamefont {S.}~\bibnamefont {Johnston}}, \bibinfo {author}
  {\bibfnamefont {T.}~\bibnamefont {Tohyama}}, \bibinfo {author} {\bibfnamefont
  {B.}~\bibnamefont {Moritz}}, \ and\ \bibinfo {author} {\bibfnamefont {T.~P.}\
  \bibnamefont {Devereaux}},\ }\bibfield  {title} {\enquote {\bibinfo {title}
  {{Persistent spin excitations in doped antiferromagnets revealed by resonant
  inelastic light scattering}},}\ }\href {http://dx.doi.org/10.1038/ncomms4314}
  {\bibfield  {journal} {\bibinfo  {journal} {Nat. Commun.}\ }\textbf {\bibinfo
  {volume} {5}},\ \bibinfo {pages} {3314} (\bibinfo {year} {2014})}\BibitemShut
  {NoStop}%
\bibitem [{\citenamefont {Peng}\ \emph {et~al.}(2018)\citenamefont {Peng},
  \citenamefont {Huang}, \citenamefont {Fumagalli}, \citenamefont {Minola},
  \citenamefont {Wang}, \citenamefont {Sun}, \citenamefont {Ding},
  \citenamefont {Kummer}, \citenamefont {Zhou}, \citenamefont {Brookes},
  \citenamefont {Moritz}, \citenamefont {Braicovich}, \citenamefont
  {Devereaux},\ and\ \citenamefont {Ghiringhelli}}]{PengPhysRevB2018}%
  \BibitemOpen
  \bibfield  {author} {\bibinfo {author} {\bibfnamefont {Y.~Y.}\ \bibnamefont
  {Peng}}, \bibinfo {author} {\bibfnamefont {E.~W.}\ \bibnamefont {Huang}},
  \bibinfo {author} {\bibfnamefont {R.}~\bibnamefont {Fumagalli}}, \bibinfo
  {author} {\bibfnamefont {M.}~\bibnamefont {Minola}}, \bibinfo {author}
  {\bibfnamefont {Y.}~\bibnamefont {Wang}}, \bibinfo {author} {\bibfnamefont
  {X.}~\bibnamefont {Sun}}, \bibinfo {author} {\bibfnamefont {Y.}~\bibnamefont
  {Ding}}, \bibinfo {author} {\bibfnamefont {K.}~\bibnamefont {Kummer}},
  \bibinfo {author} {\bibfnamefont {X.~J.}\ \bibnamefont {Zhou}}, \bibinfo
  {author} {\bibfnamefont {N.~B.}\ \bibnamefont {Brookes}}, \bibinfo {author}
  {\bibfnamefont {B.}~\bibnamefont {Moritz}}, \bibinfo {author} {\bibfnamefont
  {L.}~\bibnamefont {Braicovich}}, \bibinfo {author} {\bibfnamefont {T.~P.}\
  \bibnamefont {Devereaux}}, \ and\ \bibinfo {author} {\bibfnamefont
  {G.}~\bibnamefont {Ghiringhelli}},\ }\bibfield  {title} {\enquote {\bibinfo
  {title} {{Dispersion, damping, and intensity of spin excitations in the
  monolayer $\mathrm{(Bi,Pb)_2(Sr,La)_2CuO_{6+\delta}}$ cuprate superconductor
  family}},}\ }\href {http://dx.doi.org/10.1103/physrevb.98.144507} {\bibfield
  {journal} {\bibinfo  {journal} {Phys. Rev. B}\ }\textbf {\bibinfo {volume}
  {98}} (\bibinfo {year} {2018})}\BibitemShut {NoStop}%
\bibitem [{\citenamefont {Ishii}\ \emph {et~al.}(2017)\citenamefont {Ishii},
  \citenamefont {Tohyama}, \citenamefont {Asano}, \citenamefont {Sato},
  \citenamefont {Fujita}, \citenamefont {Wakimoto}, \citenamefont {Tustsui},
  \citenamefont {Sota}, \citenamefont {Miyawaki}, \citenamefont {Niwa},
  \citenamefont {Harada}, \citenamefont {Pelliciari}, \citenamefont {Huang},
  \citenamefont {Schmitt}, \citenamefont {Yamamoto},\ and\ \citenamefont
  {Mizuki}}]{IshiiPhysRevB2017}%
  \BibitemOpen
  \bibfield  {author} {\bibinfo {author} {\bibfnamefont {K.}~\bibnamefont
  {Ishii}}, \bibinfo {author} {\bibfnamefont {T.}~\bibnamefont {Tohyama}},
  \bibinfo {author} {\bibfnamefont {S.}~\bibnamefont {Asano}}, \bibinfo
  {author} {\bibfnamefont {K.}~\bibnamefont {Sato}}, \bibinfo {author}
  {\bibfnamefont {M.}~\bibnamefont {Fujita}}, \bibinfo {author} {\bibfnamefont
  {S.}~\bibnamefont {Wakimoto}}, \bibinfo {author} {\bibfnamefont
  {K.}~\bibnamefont {Tustsui}}, \bibinfo {author} {\bibfnamefont
  {S.}~\bibnamefont {Sota}}, \bibinfo {author} {\bibfnamefont {J.}~\bibnamefont
  {Miyawaki}}, \bibinfo {author} {\bibfnamefont {H.}~\bibnamefont {Niwa}},
  \bibinfo {author} {\bibfnamefont {Y.}~\bibnamefont {Harada}}, \bibinfo
  {author} {\bibfnamefont {J.}~\bibnamefont {Pelliciari}}, \bibinfo {author}
  {\bibfnamefont {Y.}~\bibnamefont {Huang}}, \bibinfo {author} {\bibfnamefont
  {T.}~\bibnamefont {Schmitt}}, \bibinfo {author} {\bibfnamefont
  {Y.}~\bibnamefont {Yamamoto}}, \ and\ \bibinfo {author} {\bibfnamefont
  {J.}~\bibnamefont {Mizuki}},\ }\bibfield  {title} {\enquote {\bibinfo {title}
  {{Observation of momentum-dependent charge excitations in hole-doped cuprates
  using resonant inelastic $x$-ray scattering at the oxygen $K$ edge}},}\
  }\href {http://dx.doi.org/10.1103/physrevb.96.115148} {\bibfield  {journal}
  {\bibinfo  {journal} {Phys. Rev. B}\ }\textbf {\bibinfo {volume} {96}},\
  \bibinfo {pages} {115148} (\bibinfo {year} {2017})}\BibitemShut {NoStop}%
\bibitem [{\citenamefont {Hepting}\ \emph {et~al.}(2018)\citenamefont
  {Hepting}, \citenamefont {Chaix}, \citenamefont {Huang}, \citenamefont
  {Fumagalli}, \citenamefont {Peng}, \citenamefont {Moritz}, \citenamefont
  {Kummer}, \citenamefont {Brookes}, \citenamefont {Lee}, \citenamefont
  {Hashimoto}, \citenamefont {Sarkar}, \citenamefont {He}, \citenamefont
  {Rotundu}, \citenamefont {Lee}, \citenamefont {Greene}, \citenamefont
  {Braicovich}, \citenamefont {Ghiringhelli}, \citenamefont {Shen},
  \citenamefont {Devereaux},\ and\ \citenamefont {Lee}}]{HeptingNature2018}%
  \BibitemOpen
  \bibfield  {author} {\bibinfo {author} {\bibfnamefont {M.}~\bibnamefont
  {Hepting}}, \bibinfo {author} {\bibfnamefont {L.}~\bibnamefont {Chaix}},
  \bibinfo {author} {\bibfnamefont {E.~W.}\ \bibnamefont {Huang}}, \bibinfo
  {author} {\bibfnamefont {R.}~\bibnamefont {Fumagalli}}, \bibinfo {author}
  {\bibfnamefont {Y.~Y.}\ \bibnamefont {Peng}}, \bibinfo {author}
  {\bibfnamefont {B.}~\bibnamefont {Moritz}}, \bibinfo {author} {\bibfnamefont
  {K.}~\bibnamefont {Kummer}}, \bibinfo {author} {\bibfnamefont {N.~B.}\
  \bibnamefont {Brookes}}, \bibinfo {author} {\bibfnamefont {W.~C.}\
  \bibnamefont {Lee}}, \bibinfo {author} {\bibfnamefont {M.}~\bibnamefont
  {Hashimoto}}, \bibinfo {author} {\bibfnamefont {T.}~\bibnamefont {Sarkar}},
  \bibinfo {author} {\bibfnamefont {J.-F.}\ \bibnamefont {He}}, \bibinfo
  {author} {\bibfnamefont {C.~R.}\ \bibnamefont {Rotundu}}, \bibinfo {author}
  {\bibfnamefont {Y.~S.}\ \bibnamefont {Lee}}, \bibinfo {author} {\bibfnamefont
  {R.~L.}\ \bibnamefont {Greene}}, \bibinfo {author} {\bibfnamefont
  {L.}~\bibnamefont {Braicovich}}, \bibinfo {author} {\bibfnamefont
  {G.}~\bibnamefont {Ghiringhelli}}, \bibinfo {author} {\bibfnamefont {Z.~X.}\
  \bibnamefont {Shen}}, \bibinfo {author} {\bibfnamefont {T.~P.}\ \bibnamefont
  {Devereaux}}, \ and\ \bibinfo {author} {\bibfnamefont {W.~S.}\ \bibnamefont
  {Lee}},\ }\bibfield  {title} {\enquote {\bibinfo {title} {{Three-dimensional
  collective charge excitations in electron-doped copper oxide
  superconductors}},}\ }\href {http://dx.doi.org/10.1038/s41586-018-0648-3}
  {\bibfield  {journal} {\bibinfo  {journal} {Nature}\ }\textbf {\bibinfo
  {volume} {563}},\ \bibinfo {pages} {374} (\bibinfo {year}
  {2018})}\BibitemShut {NoStop}%
\bibitem [{\citenamefont {Ishii}\ \emph {et~al.}(2019)\citenamefont {Ishii},
  \citenamefont {Kurooka}, \citenamefont {Shimizu}, \citenamefont {Fujita},
  \citenamefont {Yamada},\ and\ \citenamefont
  {Mizuki}}]{IshiiJPhysSocJapan2019}%
  \BibitemOpen
  \bibfield  {author} {\bibinfo {author} {\bibfnamefont {K.}~\bibnamefont
  {Ishii}}, \bibinfo {author} {\bibfnamefont {M.}~\bibnamefont {Kurooka}},
  \bibinfo {author} {\bibfnamefont {Y.}~\bibnamefont {Shimizu}}, \bibinfo
  {author} {\bibfnamefont {M.}~\bibnamefont {Fujita}}, \bibinfo {author}
  {\bibfnamefont {K.}~\bibnamefont {Yamada}}, \ and\ \bibinfo {author}
  {\bibfnamefont {J.}~\bibnamefont {Mizuki}},\ }\bibfield  {title} {\enquote
  {\bibinfo {title} {{Charge Excitations in
  $\mathrm{Nd_{2-\mathit{x}}Ce_\mathit{x}CuO_4}$ Observed with Resonant
  Inelastic $X$-ray Scattering: Comparison of Cu $K$-edge with Cu
  $L_3$-edge}},}\ }\href {\doibase 10.7566/JPSJ.88.075001} {\bibfield
  {journal} {\bibinfo  {journal} {J. Phys. Soc. Japan}\ }\textbf {\bibinfo
  {volume} {88}},\ \bibinfo {pages} {075001} (\bibinfo {year}
  {2019})}\BibitemShut {NoStop}%
\bibitem [{\citenamefont {Lin}\ \emph {et~al.}(2020)\citenamefont {Lin},
  \citenamefont {Yuan}, \citenamefont {Jin}, \citenamefont {Yin}, \citenamefont
  {Li}, \citenamefont {Zhou}, \citenamefont {Lu}, \citenamefont {Dantz},
  \citenamefont {Schmitt}, \citenamefont {Ding}, \citenamefont {Guo},
  \citenamefont {P.~M.~Dean},\ and\ \citenamefont
  {Liu}}]{LinNPJQuantMater2020}%
  \BibitemOpen
  \bibfield  {author} {\bibinfo {author} {\bibfnamefont {J.}~\bibnamefont
  {Lin}}, \bibinfo {author} {\bibfnamefont {J.}~\bibnamefont {Yuan}}, \bibinfo
  {author} {\bibfnamefont {K.}~\bibnamefont {Jin}}, \bibinfo {author}
  {\bibfnamefont {Z.}~\bibnamefont {Yin}}, \bibinfo {author} {\bibfnamefont
  {Gang}\ \bibnamefont {Li}}, \bibinfo {author} {\bibfnamefont {K.-J.}\
  \bibnamefont {Zhou}}, \bibinfo {author} {\bibfnamefont {X.}~\bibnamefont
  {Lu}}, \bibinfo {author} {\bibfnamefont {M.}~\bibnamefont {Dantz}}, \bibinfo
  {author} {\bibfnamefont {T.}~\bibnamefont {Schmitt}}, \bibinfo {author}
  {\bibfnamefont {H.}~\bibnamefont {Ding}}, \bibinfo {author} {\bibfnamefont
  {H.}~\bibnamefont {Guo}}, \bibinfo {author} {\bibfnamefont {M.}~\bibnamefont
  {P.~M.~Dean}}, \ and\ \bibinfo {author} {\bibfnamefont {X.}~\bibnamefont
  {Liu}},\ }\bibfield  {title} {\enquote {\bibinfo {title} {{Doping evolution
  of the charge excitations and electron correlations in electron-doped
  superconducting $\mathrm{La_{2-\mathit{x}}Ce_{\mathit{x}}CuO_4}$}},}\ }\href
  {\doibase 10.1038/s41535-019-0205-9} {\bibfield  {journal} {\bibinfo
  {journal} {npj Quant. Mater.}\ }\textbf {\bibinfo {volume} {5}},\ \bibinfo
  {pages} {4} (\bibinfo {year} {2020})}\BibitemShut {NoStop}%
\bibitem [{\citenamefont {Singh}\ \emph {et~al.}(2020)\citenamefont {Singh},
  \citenamefont {Huang}, \citenamefont {Lane}, \citenamefont {Li},
  \citenamefont {Okamoto}, \citenamefont {Komiya}, \citenamefont {Markiewicz},
  \citenamefont {Bansil}, \citenamefont {Fujimori}, \citenamefont {Chen},\ and\
  \citenamefont {Huang}}]{SinghArXiV2020}%
  \BibitemOpen
  \bibfield  {author} {\bibinfo {author} {\bibfnamefont {A.}~\bibnamefont
  {Singh}}, \bibinfo {author} {\bibfnamefont {H.~Y.}\ \bibnamefont {Huang}},
  \bibinfo {author} {\bibfnamefont {Christopher}\ \bibnamefont {Lane}},
  \bibinfo {author} {\bibfnamefont {J.~H.}\ \bibnamefont {Li}}, \bibinfo
  {author} {\bibfnamefont {J.}~\bibnamefont {Okamoto}}, \bibinfo {author}
  {\bibfnamefont {S.}~\bibnamefont {Komiya}}, \bibinfo {author} {\bibfnamefont
  {Robert~S.}\ \bibnamefont {Markiewicz}}, \bibinfo {author} {\bibfnamefont
  {Arun}\ \bibnamefont {Bansil}}, \bibinfo {author} {\bibfnamefont
  {A.}~\bibnamefont {Fujimori}}, \bibinfo {author} {\bibfnamefont {C.~T.}\
  \bibnamefont {Chen}}, \ and\ \bibinfo {author} {\bibfnamefont {D.~J.}\
  \bibnamefont {Huang}},\ }\href@noop {} {\enquote {\bibinfo {title} {{Acoustic
  plasmons and conducting carriers in hole-doped cuprate superconductors}},}\ }
  (\bibinfo {year} {2020}),\ \Eprint {http://arxiv.org/abs/2006.13424}
  {arXiv:2006.13424} \BibitemShut {NoStop}%
\bibitem [{\citenamefont {Nag}\ \emph {et~al.}(2020)\citenamefont {Nag},
  \citenamefont {Zhu}, \citenamefont {Bejas}, \citenamefont {Li}, \citenamefont
  {Robarts}, \citenamefont {Yamase}, \citenamefont {Petsch}, \citenamefont
  {Song}, \citenamefont {Eisaki}, \citenamefont {Walters}, \citenamefont
  {Garc\'{\i}a-Fern\'andez}, \citenamefont {Greco}, \citenamefont {Hayden},\
  and\ \citenamefont {Zhou}}]{NagPhysRevLett2020}%
  \BibitemOpen
  \bibfield  {author} {\bibinfo {author} {\bibfnamefont {A.}~\bibnamefont
  {Nag}}, \bibinfo {author} {\bibfnamefont {M.}~\bibnamefont {Zhu}}, \bibinfo
  {author} {\bibfnamefont {M.}~\bibnamefont {Bejas}}, \bibinfo {author}
  {\bibfnamefont {J.}~\bibnamefont {Li}}, \bibinfo {author} {\bibfnamefont
  {H.~C.}\ \bibnamefont {Robarts}}, \bibinfo {author} {\bibfnamefont
  {H.}~\bibnamefont {Yamase}}, \bibinfo {author} {\bibfnamefont {A.~N.}\
  \bibnamefont {Petsch}}, \bibinfo {author} {\bibfnamefont {D.}~\bibnamefont
  {Song}}, \bibinfo {author} {\bibfnamefont {H.}~\bibnamefont {Eisaki}},
  \bibinfo {author} {\bibfnamefont {A.~C.}\ \bibnamefont {Walters}}, \bibinfo
  {author} {\bibfnamefont {M.}~\bibnamefont {Garc\'{\i}a-Fern\'andez}},
  \bibinfo {author} {\bibfnamefont {A.}~\bibnamefont {Greco}}, \bibinfo
  {author} {\bibfnamefont {S.~M.}\ \bibnamefont {Hayden}}, \ and\ \bibinfo
  {author} {\bibfnamefont {K.-J.}\ \bibnamefont {Zhou}},\ }\bibfield  {title}
  {\enquote {\bibinfo {title} {{Detection of Acoustic Plasmons in Hole-Doped
  Lanthanum and Bismuth Cuprate Superconductors Using Resonant Inelastic X-Ray
  Scattering}},}\ }\href {https://doi.org/10.1103/PhysRevLett.125.257002}
  {\bibfield  {journal} {\bibinfo  {journal} {Phys. Rev. Lett.}\ }\textbf
  {\bibinfo {volume} {125}},\ \bibinfo {pages} {257002} (\bibinfo {year}
  {2020})}\BibitemShut {NoStop}%
\bibitem [{\citenamefont {Ovchinnikov}\ and\ \citenamefont
  {Val'kov}(2004)}]{OvchinnikovBook2004}%
  \BibitemOpen
  \bibfield  {author} {\bibinfo {author} {\bibfnamefont {S.~G.}\ \bibnamefont
  {Ovchinnikov}}\ and\ \bibinfo {author} {\bibfnamefont {V.~V.}\ \bibnamefont
  {Val'kov}},\ }\href@noop {} {\emph {\bibinfo {title} {{Hubbard operators in
  the theory of strongly correlated electrons}}}}\ (\bibinfo  {publisher}
  {Imperial College Press, London},\ \bibinfo {year} {2004})\BibitemShut
  {NoStop}%
\bibitem [{\citenamefont {Fidrysiak}\ and\ \citenamefont
  {Spa{\l}ek}(2020)}]{FidrysiakPhysRevB2020}%
  \BibitemOpen
  \bibfield  {author} {\bibinfo {author} {\bibfnamefont {M.}~\bibnamefont
  {Fidrysiak}}\ and\ \bibinfo {author} {\bibfnamefont {J.}~\bibnamefont
  {Spa{\l}ek}},\ }\bibfield  {title} {\enquote {\bibinfo {title} {{Robust spin
  and charge excitations throughout the high-$T_c$ cuprate phase diagram from
  incipient Mottness}},}\ }\href
  {http://dx.doi.org/10.1103/physrevb.102.014505} {\bibfield  {journal}
  {\bibinfo  {journal} {Phys. Rev. B}\ }\textbf {\bibinfo {volume} {102}},\
  \bibinfo {pages} {014505} (\bibinfo {year} {2020})}\BibitemShut {NoStop}%
\bibitem [{\citenamefont {Fidrysiak}\ and\ \citenamefont
  {Spa\l{}ek}(2021)}]{FidrysiakPhysRevB2021}%
  \BibitemOpen
  \bibfield  {author} {\bibinfo {author} {\bibfnamefont {M.}~\bibnamefont
  {Fidrysiak}}\ and\ \bibinfo {author} {\bibfnamefont {J.}~\bibnamefont
  {Spa\l{}ek}},\ }\bibfield  {title} {\enquote {\bibinfo {title} {{Universal
  collective modes from strong electronic correlations: Modified
  $1/{\mathcal{N}}_{f}$ theory with application to high-${T}_{c}$ cuprates}},}\
  }\href {\doibase 10.1103/PhysRevB.103.165111} {\bibfield  {journal} {\bibinfo
   {journal} {Phys. Rev. B}\ }\textbf {\bibinfo {volume} {103}},\ \bibinfo
  {pages} {165111} (\bibinfo {year} {2021})}\BibitemShut {NoStop}%
\bibitem [{\citenamefont {Fidrysiak}\ and\ \citenamefont
  {Spa{\l}ek}()}]{FidrysiakArXiV2021}%
  \BibitemOpen
  \bibfield  {author} {\bibinfo {author} {\bibfnamefont {M.}~\bibnamefont
  {Fidrysiak}}\ and\ \bibinfo {author} {\bibfnamefont {J.}~\bibnamefont
  {Spa{\l}ek}},\ }\bibfield  {title} {\enquote {\bibinfo {title} {{A unified
  theory of spin and charge excitations in high-$T_c$ cuprates: Quantitative
  comparison with experiment and interpretation}},}\ }\href@noop {} {\bibinfo
  {journal} {\href{https://arxiv.org/abs/2104.12812}{arXiv:2104.12812}}\
  }\BibitemShut {NoStop}%
\bibitem [{\citenamefont {Greco}\ \emph {et~al.}(2016)\citenamefont {Greco},
  \citenamefont {Yamase},\ and\ \citenamefont {Bejas}}]{GrecoPhysRevB2016}%
  \BibitemOpen
\bibfield  {journal} {  }\bibfield  {author} {\bibinfo {author} {\bibfnamefont
  {A.}~\bibnamefont {Greco}}, \bibinfo {author} {\bibfnamefont
  {H.}~\bibnamefont {Yamase}}, \ and\ \bibinfo {author} {\bibfnamefont
  {M.}~\bibnamefont {Bejas}},\ }\bibfield  {title} {\enquote {\bibinfo {title}
  {{Plasmon excitations in layered high-$T_c$ cuprates}},}\ }\href
  {http://dx.doi.org/10.1103/physrevb.94.075139} {\bibfield  {journal}
  {\bibinfo  {journal} {Phys. Rev. B}\ }\textbf {\bibinfo {volume} {94}},\
  \bibinfo {pages} {075139} (\bibinfo {year} {2016})}\BibitemShut {NoStop}%
\bibitem [{\citenamefont {Greco}\ \emph {et~al.}(2017)\citenamefont {Greco},
  \citenamefont {Yamase},\ and\ \citenamefont {Bejas}}]{GrecoJPSJ2017}%
  \BibitemOpen
  \bibfield  {author} {\bibinfo {author} {\bibfnamefont {A.}~\bibnamefont
  {Greco}}, \bibinfo {author} {\bibfnamefont {H.}~\bibnamefont {Yamase}}, \
  and\ \bibinfo {author} {\bibfnamefont {M.}~\bibnamefont {Bejas}},\ }\bibfield
   {title} {\enquote {\bibinfo {title} {{Charge-Density-Excitation Spectrum in
  the $t$-$t^\prime$-$J$-$V$ Model}},}\ }\href
  {https://doi.org/10.7566/JPSJ.86.034706} {\bibfield  {journal} {\bibinfo
  {journal} {J. Phys. Soc. Japan}\ }\textbf {\bibinfo {volume} {86}},\ \bibinfo
  {pages} {034706} (\bibinfo {year} {2017})}\BibitemShut {NoStop}%
\bibitem [{\citenamefont {Greco}\ \emph {et~al.}(2020)\citenamefont {Greco},
  \citenamefont {Yamase},\ and\ \citenamefont {Bejas}}]{GrecoPhysRevB2020}%
  \BibitemOpen
  \bibfield  {author} {\bibinfo {author} {\bibfnamefont {A.}~\bibnamefont
  {Greco}}, \bibinfo {author} {\bibfnamefont {H.}~\bibnamefont {Yamase}}, \
  and\ \bibinfo {author} {\bibfnamefont {M.}~\bibnamefont {Bejas}},\ }\bibfield
   {title} {\enquote {\bibinfo {title} {{Close inspection of plasmon
  excitations in cuprate superconductors}},}\ }\href
  {https://doi.org/10.1103/PhysRevB.102.024509} {\bibfield  {journal} {\bibinfo
   {journal} {Phys. Rev. B}\ }\textbf {\bibinfo {volume} {102}},\ \bibinfo
  {pages} {024509} (\bibinfo {year} {2020})}\BibitemShut {NoStop}%
\bibitem [{\citenamefont {Chao}\ \emph {et~al.}(1978)\citenamefont {Chao},
  \citenamefont {Spa{\l}ek},\ and\ \citenamefont
  {Ole\'{s}}}]{ChaoPhysRevB1978}%
  \BibitemOpen
  \bibfield  {author} {\bibinfo {author} {\bibfnamefont {K.~A.}\ \bibnamefont
  {Chao}}, \bibinfo {author} {\bibfnamefont {J.}~\bibnamefont {Spa{\l}ek}}, \
  and\ \bibinfo {author} {\bibfnamefont {A.~M.}\ \bibnamefont {Ole\'{s}}},\
  }\bibfield  {title} {\enquote {\bibinfo {title} {{Canonical perturbation
  expansion of the Hubbard model}},}\ }\href
  {http://dx.doi.org/10.1103/physrevb.18.3453} {\bibfield  {journal} {\bibinfo
  {journal} {Phys. Rev. B}\ }\textbf {\bibinfo {volume} {18}},\ \bibinfo
  {pages} {3453} (\bibinfo {year} {1978})}\BibitemShut {NoStop}%
\bibitem [{\citenamefont {Spa{\l}ek}\ \emph {et~al.}(1981)\citenamefont
  {Spa{\l}ek}, \citenamefont {Ole\'{s}},\ and\ \citenamefont
  {Chao}}]{SpalekPhysstatSol1981}%
  \BibitemOpen
  \bibfield  {author} {\bibinfo {author} {\bibfnamefont {J.}~\bibnamefont
  {Spa{\l}ek}}, \bibinfo {author} {\bibfnamefont {A.~M.}\ \bibnamefont
  {Ole\'{s}}}, \ and\ \bibinfo {author} {\bibfnamefont {K.~A.}\ \bibnamefont
  {Chao}},\ }\bibfield  {title} {\enquote {\bibinfo {title} {{Magnetic Phases
  of Strongly Correlated Electrons in a Nearly Half-Filled Narrow Band}},}\
  }\href {\doibase 10.1002/pssb.2221080206} {\bibfield  {journal} {\bibinfo
  {journal} {phys. stat. sol. (b)}\ }\textbf {\bibinfo {volume} {108}},\
  \bibinfo {pages} {329} (\bibinfo {year} {1981})}\BibitemShut {NoStop}%
\bibitem [{\citenamefont {Eisenberg}\ \emph {et~al.}(2002)\citenamefont
  {Eisenberg}, \citenamefont {Berkovits}, \citenamefont {Huse},\ and\
  \citenamefont {Altshuler}}]{EisenbergPhysRevB2002}%
  \BibitemOpen
  \bibfield  {author} {\bibinfo {author} {\bibfnamefont {E.}~\bibnamefont
  {Eisenberg}}, \bibinfo {author} {\bibfnamefont {R.}~\bibnamefont
  {Berkovits}}, \bibinfo {author} {\bibfnamefont {David~A.}\ \bibnamefont
  {Huse}}, \ and\ \bibinfo {author} {\bibfnamefont {B.~L.}\ \bibnamefont
  {Altshuler}},\ }\bibfield  {title} {\enquote {\bibinfo {title} {{Breakdown of
  the Nagaoka phase in the two-dimensional $t$-$J$ model}},}\ }\href {\doibase
  10.1103/PhysRevB.65.134437} {\bibfield  {journal} {\bibinfo  {journal} {Phys.
  Rev. B}\ }\textbf {\bibinfo {volume} {65}},\ \bibinfo {pages} {134437}
  (\bibinfo {year} {2002})}\BibitemShut {NoStop}%
\bibitem [{\citenamefont {Liu}\ \emph {et~al.}(2012)\citenamefont {Liu},
  \citenamefont {Yao}, \citenamefont {Berg}, \citenamefont {White},\ and\
  \citenamefont {Kivelson}}]{LiuPhysRevLett2012}%
  \BibitemOpen
  \bibfield  {author} {\bibinfo {author} {\bibfnamefont {L.}~\bibnamefont
  {Liu}}, \bibinfo {author} {\bibfnamefont {H.}~\bibnamefont {Yao}}, \bibinfo
  {author} {\bibfnamefont {E.}~\bibnamefont {Berg}}, \bibinfo {author}
  {\bibfnamefont {S.~R.}\ \bibnamefont {White}}, \ and\ \bibinfo {author}
  {\bibfnamefont {S.~A.}\ \bibnamefont {Kivelson}},\ }\bibfield  {title}
  {\enquote {\bibinfo {title} {{Phases of the Infinite $U$ Hubbard Model on
  Square Lattices}},}\ }\href {\doibase 10.1103/PhysRevLett.108.126406}
  {\bibfield  {journal} {\bibinfo  {journal} {Phys. Rev. Lett.}\ }\textbf
  {\bibinfo {volume} {108}},\ \bibinfo {pages} {126406} (\bibinfo {year}
  {2012})}\BibitemShut {NoStop}%
\bibitem [{\citenamefont {Blesio}\ \emph {et~al.}(2019)\citenamefont {Blesio},
  \citenamefont {Gonzalez},\ and\ \citenamefont
  {Lisandrini}}]{BlesioPhysRevB2019}%
  \BibitemOpen
  \bibfield  {author} {\bibinfo {author} {\bibfnamefont {G.~G.}\ \bibnamefont
  {Blesio}}, \bibinfo {author} {\bibfnamefont {M.~G.}\ \bibnamefont
  {Gonzalez}}, \ and\ \bibinfo {author} {\bibfnamefont {F.~T.}\ \bibnamefont
  {Lisandrini}},\ }\bibfield  {title} {\enquote {\bibinfo {title} {{Magnetic
  phase diagram of the infinite-$U$ Hubbard model with nearest- and
  next-nearest-neighbor hoppings}},}\ }\href {\doibase
  10.1103/PhysRevB.99.174411} {\bibfield  {journal} {\bibinfo  {journal} {Phys.
  Rev. B}\ }\textbf {\bibinfo {volume} {99}},\ \bibinfo {pages} {174411}
  (\bibinfo {year} {2019})}\BibitemShut {NoStop}%
\bibitem [{\citenamefont {B\"{u}nemann}\ \emph {et~al.}(2012)\citenamefont
  {B\"{u}nemann}, \citenamefont {Schickling},\ and\ \citenamefont
  {Gebhard}}]{BunemannEPL2012}%
  \BibitemOpen
  \bibfield  {author} {\bibinfo {author} {\bibfnamefont {J.}~\bibnamefont
  {B\"{u}nemann}}, \bibinfo {author} {\bibfnamefont {T.}~\bibnamefont
  {Schickling}}, \ and\ \bibinfo {author} {\bibfnamefont {F.}~\bibnamefont
  {Gebhard}},\ }\bibfield  {title} {\enquote {\bibinfo {title} {{Variational
  study of Fermi surface deformations in Hubbard models}},}\ }\href {\doibase
  10.1209/0295-5075/98/27006} {\bibfield  {journal} {\bibinfo  {journal} {EPL}\
  }\textbf {\bibinfo {volume} {98}},\ \bibinfo {pages} {27006} (\bibinfo {year}
  {2012})}\BibitemShut {NoStop}%
\bibitem [{\citenamefont {Kaczmarczyk}\ \emph {et~al.}(2014)\citenamefont
  {Kaczmarczyk}, \citenamefont {B\"{u}nemann},\ and\ \citenamefont
  {Spa{\l}ek}}]{KaczmarczykNewJPhys2014}%
  \BibitemOpen
  \bibfield  {author} {\bibinfo {author} {\bibfnamefont {J.}~\bibnamefont
  {Kaczmarczyk}}, \bibinfo {author} {\bibfnamefont {J.}~\bibnamefont
  {B\"{u}nemann}}, \ and\ \bibinfo {author} {\bibfnamefont {J.}~\bibnamefont
  {Spa{\l}ek}},\ }\bibfield  {title} {\enquote {\bibinfo {title}
  {{High-temperature superconductivity in the two-dimensional $t$-$J$ model:
  Gutzwiller wavefunction solution}},}\ }\href
  {http://dx.doi.org/10.1088/1367-2630/16/7/073018} {\bibfield  {journal}
  {\bibinfo  {journal} {New J. Phys.}\ }\textbf {\bibinfo {volume} {16}},\
  \bibinfo {pages} {073018} (\bibinfo {year} {2014})}\BibitemShut {NoStop}%
\bibitem [{\citenamefont {Kresin}\ and\ \citenamefont
  {Morawitz}(1988)}]{KresinPhysRevB1988}%
  \BibitemOpen
  \bibfield  {author} {\bibinfo {author} {\bibfnamefont {V.~Z.}\ \bibnamefont
  {Kresin}}\ and\ \bibinfo {author} {\bibfnamefont {H.}~\bibnamefont
  {Morawitz}},\ }\bibfield  {title} {\enquote {\bibinfo {title} {{Layer
  plasmons and high-${T}_{c}$ superconductivity}},}\ }\href {\doibase
  10.1103/PhysRevB.37.7854} {\bibfield  {journal} {\bibinfo  {journal} {Phys.
  Rev. B}\ }\textbf {\bibinfo {volume} {37}},\ \bibinfo {pages} {7854}
  (\bibinfo {year} {1988})}\BibitemShut {NoStop}%
\bibitem [{\citenamefont {Aristov}\ and\ \citenamefont
  {Khaliullin}(2006)}]{AristovPhysRevB2006}%
  \BibitemOpen
  \bibfield  {author} {\bibinfo {author} {\bibfnamefont {D.~N.}\ \bibnamefont
  {Aristov}}\ and\ \bibinfo {author} {\bibfnamefont {G.}~\bibnamefont
  {Khaliullin}},\ }\bibfield  {title} {\enquote {\bibinfo {title} {{Charge
  susceptibility in the $t$-$J$ model}},}\ }\href {\doibase
  10.1103/PhysRevB.74.045124} {\bibfield  {journal} {\bibinfo  {journal} {Phys.
  Rev. B}\ }\textbf {\bibinfo {volume} {74}},\ \bibinfo {pages} {045124}
  (\bibinfo {year} {2006})}\BibitemShut {NoStop}%
\bibitem [{\citenamefont {Dahm}\ \emph {et~al.}(2009)\citenamefont {Dahm},
  \citenamefont {Hinkov}, \citenamefont {Borisenko}, \citenamefont {Kordyuk},
  \citenamefont {Zabolotnyy}, \citenamefont {Fink}, \citenamefont
  {B\"{u}chner}, \citenamefont {Scalapino}, \citenamefont {Hanke},\ and\
  \citenamefont {Keimer}}]{DahmNatPhys2009}%
  \BibitemOpen
  \bibfield  {author} {\bibinfo {author} {\bibfnamefont {T.}~\bibnamefont
  {Dahm}}, \bibinfo {author} {\bibfnamefont {V.}~\bibnamefont {Hinkov}},
  \bibinfo {author} {\bibfnamefont {S.~V.}\ \bibnamefont {Borisenko}}, \bibinfo
  {author} {\bibfnamefont {A.~A.}\ \bibnamefont {Kordyuk}}, \bibinfo {author}
  {\bibfnamefont {V.~B.}\ \bibnamefont {Zabolotnyy}}, \bibinfo {author}
  {\bibfnamefont {J.}~\bibnamefont {Fink}}, \bibinfo {author} {\bibfnamefont
  {B.}~\bibnamefont {B\"{u}chner}}, \bibinfo {author} {\bibfnamefont {D.~J.}\
  \bibnamefont {Scalapino}}, \bibinfo {author} {\bibfnamefont {W.}~\bibnamefont
  {Hanke}}, \ and\ \bibinfo {author} {\bibfnamefont {B.}~\bibnamefont
  {Keimer}},\ }\bibfield  {title} {\enquote {\bibinfo {title} {{Strength of the
  spin-fluctuation-mediated pairing interaction in a high-temperature
  superconductor}},}\ }\href {\doibase 10.1038/nphys1180} {\bibfield  {journal}
  {\bibinfo  {journal} {Nat. Phys.}\ }\textbf {\bibinfo {volume} {5}},\
  \bibinfo {pages} {217} (\bibinfo {year} {2009})}\BibitemShut {NoStop}%
\end{thebibliography}
\end{document}